\documentclass[journal,twoside,web]{ieeecolor}
\usepackage{tmp_st}
\usepackage{cite}
\usepackage{amsmath,amssymb,amsfonts}
\usepackage{algorithmic}
\usepackage{graphicx}
\usepackage{float}
\usepackage{adjustbox}
\usepackage{caption}
\usepackage{stfloats}
\usepackage[justification=justified]{caption}
\usepackage{textcomp}
\usepackage[utf8]{inputenc}
\usepackage{longtable}
\usepackage{url}
\usepackage{xr}
\usepackage{multirow}
\usepackage{booktabs}
\usepackage[switch,columnwise]{lineno}
\usepackage{amsmath,amssymb,amsfonts}
\usepackage{float}
\usepackage{adjustbox}
\usepackage{stfloats}
\usepackage[justification=justified]{caption}
\usepackage{textcomp}
\usepackage{longtable}
\usepackage{url}
\usepackage{xr}
\usepackage{multirow}
\usepackage{xcolor}
\usepackage[colorlinks]{hyperref}
\usepackage{nicefrac}       
\usepackage{microtype}      
\usepackage{booktabs}
\usepackage{multirow}
\usepackage{siunitx}
\usepackage{graphicx}
\usepackage{array}

\DeclareUnicodeCharacter{2081}{\ensuremath{{}_1}}

\definecolor{bondiblue}{rgb}{0.0, 0.58, 0.71}
\definecolor{brightcerulean}{rgb}{0.11, 0.62, 0.74}

\usepackage{chngcntr}
\counterwithin*{subsubsection}{subsection}
\renewcommand{\thesubsubsection}{\thesubsection.\alph{subsubsection}}

\def\SP#1{\textsuperscript{#1}}
\def\SB#1{\textsubscript{#1}}
\def\BibTeX{{\rm B\kern-.05em{\sc i\kern-.025em b}\kern-.08em
    T\kern-.1667em\lower.7ex\hbox{E}\kern-.125emX}}
\begin{document}
\title{Learning Deep MRI Reconstruction Models from Scratch in Low-Data Regimes}
\author{Salman Ul Hassan Dar, \c{S}aban \"{O}zt\"{u}rk, Muzaffer {\"O}zbey, and Tolga \c{C}ukur$^*$
\thanks{This work was supported in part by a TUBA GEBIP 2015 fellowship, by a BAGEP 2017 fellowship, and by a TUBITAK 121E488 grant awarded to T. \c{C}ukur.}
\thanks{S. UH. Dar,  \c{S}. \"{O}zt\"{u}rk, M. Özbey, and T. Çukur are with the Department of Electrical and Electronics Engineering, and the National Magnetic Resonance Research Center, Bilkent University, Ankara, Turkey (e-mails: \{salman,muzaffer,cukur\}@ee.bilkent.edu.tr, saban.ozturk@amasya.edu.tr). Ş. Öztürk is also with the Amasya University, Amasya, Turkey. }
}

\maketitle

\begin{abstract}
Magnetic resonance imaging (MRI) is an essential diagnostic tool that suffers from prolonged scan times. Reconstruction methods can alleviate this limitation by recovering clinically usable images from accelerated acquisitions. In particular, learning-based methods promise performance leaps by employing deep neural networks as data-driven priors. A powerful approach uses scan-specific (SS) priors that leverage information regarding the underlying physical signal model for reconstruction. SS priors are learned on each individual test scan without the need for a training dataset, albeit they suffer from computationally burdening inference with nonlinear networks. An alternative approach uses scan-general (SG) priors that instead leverage information regarding the latent features of MRI images for reconstruction. SG priors are frozen at test time for efficiency, albeit they require learning from a large training dataset. Here, we introduce a novel parallel-stream fusion model (PSFNet) that synergistically fuses SS and SG priors for performant MRI reconstruction in low-data regimes, while maintaining competitive inference times to SG methods. PSFNet implements its SG prior based on a nonlinear network, yet it forms its SS prior based on a linear network to maintain efficiency. A pervasive framework for combining multiple priors in MRI reconstruction is algorithmic unrolling that uses serially alternated projections, causing error propagation under low-data regimes. To alleviate error propagation, PSFNet combines its SS and SG priors via a novel parallel-stream architecture with learnable fusion parameters. Demonstrations are performed on multi-coil brain MRI for varying amounts of training data. PSFNet outperforms SG methods in low-data regimes, and surpasses SS methods with few tens of training samples. In both supervised and unsupervised setups, PSFNet requires an order of magnitude lower samples compared to SG methods, and enables an order of magnitude faster inference compared to SS methods. Thus, the proposed model improves deep MRI reconstruction with elevated learning and computational efficiency. 
\end{abstract}


\section{Introduction}
The unparalleled soft-tissue contrast and non-invasiveness of MRI render it a preferred modality in many diagnostic applications \cite{MRI_line,SHOEIBI202385}, and downstream imaging tasks such as classification \cite{classHU2022330} and segmentation \cite{segmenFERNANDO2023450, segmZHU2023376}. However, the adverse effects of low spin polarization at mainstream field strengths on the signal-to-noise ratio make it slower against alternate modalities such as CT \cite{Grappa_net}. Since long scan durations inevitably constrain clinical utility, there is an ever-growing interest in accelerated MRI methods to improve scan efficiency. Accelerated MRI involves an ill-posed inverse problem with the aim of mapping undersampled acquisitions in k-space to high-quality images corresponding to fully-sampled acquisitions. Conventional frameworks for solving this problem rely on parallel imaging (PI) capabilities of receive coil arrays \cite{Pruessmann1999,Griswold2002}, in conjunction with hand-constructed MRI priors \cite{Lustig2007,majumdar2015improving}. A joint objective is iteratively optimized comprising a data-consistency (DC) term based on the physical signal model, and a regularization term that enforces the MRI prior \cite{Lustig2007}. The physical model constrains reconstructed data to be consistent with acquired data while considering coil sensitivities and undersampling patterns \cite{Lustig2010}. Meanwhile, the regularization term, often based on a linear transform where data are assumed to be compressible \cite{Lustig2007}, introduces suboptimality when the distribution of MRI data diverges from the hand-constructed prior.

Deep learning (DL) methods have recently been adopted as a promising framework to improve reconstruction performance \cite{ADMM-CSNET,Schlemper2017,Hammernik2017,raki,loraki}. Inspired by traditional methods, a powerful approach is based on scan-specific (SS) priors that leverage the physical signal model to learn a reconstruction specific to each test scan, i.e. undersampled k-space data from a given test subject. Similar to autocalibration procedures in PI, a first group of SS methods perform training using a fully-sampled calibration region and then exercise learned dependencies in broader k-space \cite{spark,raki,sraki,loraki}. Following the deep image prior technique, a second group of methods use unconditional CNNs as a native MRI prior \cite{Yilmaz2021,Arora2020ismrm,Darestani2021}. These CNNs map low-dimensional latent variables onto MR images, and latents and network weights are  optimized to ensure consistency to acquired data based on the physical signal model. In general, SS priors learned on each subject at test time avoid the need for separate training datasets, and promise improved reliability against atypical anatomy. However, they suffer from long inference times that can be prohibitive particularly when nonlinear networks are adopted \cite{Knoll2019inverseGANs,Konukoglu2019,Liu2020mrm}.

 A fundamental alternative is to employ scan-general (SG) priors based on deep nonlinear networks that capture latent features of MR images \cite{ADMM-CSNET,Schlemper2017,Hammernik2017,KikiNet,Mardani2019b,MoDl,Dar2017,lee2018deep,guo2021over,yiasemis2022recurrent,hou2022idpcnn,ramzi2022nc}. Numerous successful architectures have been reported including perceptrons \cite{Kwon2017}, basic convolutional neural networks (CNNs) \cite{Wang2016,ChulYe2018,Yoon2018,Hyun2018}, residual or recurrent CNNs \cite{lee2018deep,Hauptmann2018,Hosseini2020b,Conv_recur}, generative adversarial networks (GANs) \cite{Quan2018c,rgan,Chen2021,elmas2022federated,yaqub2022gan}, transformers \cite{korkmaz2022mri,guo2022reconformer} and diffusion models \cite{dar2022adaptive,peng2022towards}. Physics-guided unrolled methods have received particular attention that combine the physical signal model as in traditional frameworks and regularization via a deep network serving as an SG prior \cite{MoDl,Schlemper2017,wang2022high, supervisedunrolledGadjimuradov, supervisedunrolled9684848}. Reconstruction is achieved via serially alternated projections through the physical signal model and the SG prior \cite{Hyun2018,Hosseini2020b,Polakjointvvn2020,deepspirit, supervisedunrolled9761583}. However, under low-data regimes, the suboptimally trained SG prior introduces errors that are propagated across the unrolled architecture, compromising performance \cite{Grappa_net,tavaf2021grappa,sandino2021accelerating}. Furthermore, learning of SG priors requires large training datasets from several tens to hundreds of subjects \cite{Dar2017,KnollGeneralization,chaudhari2021}, which can limit practicality. 

Here, we propose a novel parallel-stream fusion model (PSFNet) that consolidates SS and SG priors to enable data-efficient training and computation-efficient inference in deep MRI reconstruction\footnote{see \cite{comnet2021} for a preliminary version of this work presented at ISMRM 2021.}. PSFNet leverages an SS stream to perform linear reconstruction based on the physical signal model, and an SG stream to perform nonlinear reconstruction based on a deep network. Unlike conventional unrolled methods based on serial projections, here we propose a parallel-stream architecture with learnable fusion of SS and SG priors. Fusion parameters are adapted across cascades and training iterations to emphasize task-critical information. Comprehensive experiments on brain MRI datasets are reported to demonstrate PSFNet under both supervised and unsupervised settings \cite{Tamir2019,Cole2021,yaman2020,Liu2020,wang2022}. PSFNet is compared against an unrolled SG method \cite{MoDl}, two SS methods \cite{spark,sraki_rnn}, and conventional SPIRiT reconstructions \cite{Lustig2010}. Compared to the unrolled model, PSFNet lowers training data requirements an order of magnitude. Compared to SS models, PSFNet offers significantly faster inference times. Our main contributions are summarized below: 

\begin{itemize}
\item A novel cascaded network architecture is introduced that adaptively fuses SS and SG priors across cascades and training iterations to improve learning-based MRI reconstruction in low-data regimes.

\item The SS prior facilitates learning of the SG prior with limited data, and empowers PSFNet to successfully generalize to out-of-domain samples. 

\item The SG prior improves performance by capturing nonlinear residuals, and enhances resilience against suboptimal hyperparameter selection in the SS component.

\item Parallel-stream fusion of SS and SG priors yields robust performance with limited training data in both supervised and unsupervised settings. 
\end{itemize}



\section{Theory}

\subsection{Image Reconstruction in Accelerated MRI}
MRI reconstruction is an inverse problem that aims to recover an image from a respective undersampled acquisition:
\begin{align}
    MFx = y
    \label{eq:rec1}
\end{align}
where $F$ is the Fourier transform, $M$ is the sampling mask defining acquired k-space locations, $x$ is the multi-coil image to be reconstructed and $y$ are acquired multi-coil k-space data. 
To improve problem conditioning, additional prior information regarding the expected distribution of MR images is incorporated in the form of a regularization term: 
\begin{align}
    \hat{x} = \underset{x}{\arg\min} \quad \lambda||MFx-y||_2^2 + R(x)
    \label{eq:rec_reg_1}
\end{align}
where the first term enforces DC between reconstructed and acquired k-space data, $R(x)$ reflects the MRI prior, and $\lambda$ controls the balance between the DC and regularization terms. 

The DC term can be implemented by injecting the acquired values of k-space data into the reconstruction \cite{Schlemper2017}. Thus, mapping through a DC block is given as:
\begin{align}
f_{DC}(x) = F^{-1} \Lambda F x +\dfrac{\lambda}{1+\lambda} F^{-1}y
\end{align}
where $\Lambda$ is a diagonal matrix with diagonal entries set to $\frac{1}{1+\lambda}$ at acquired k-space locations and set to 1 in unacquired locations. 

In traditional methods, the regularization term is based on a hand-constructed transform domain where data are assumed to have a sparse representation \cite{Lustig2007}. For improved conformation to the distribution of MRI data, recent frameworks instead adopt deep network models to capture either SG priors learned from a large MRI database with hundreds of subjects, or SS priors learned from individual test scans. Learning procedures for the two types of priors are discussed below. 

\textbf{\textit{SG priors}}: In MRI, SG priors are typically adopted to suppress aliasing artifacts in the zero-filled reconstruction (i.e., inverse Fourier transform) of undersampled k-space acquisitions \cite{MoDl}. A deep network model that performs de-aliasing can be learned from a large training dataset of undersampled and corresponding fully-sampled k-space acquisitions, and then employed to implement $R(.)$ in Eq. \ref{eq:rec_reg_1} during inference. The regularization term based on an SG prior is given as:
\begin{align}
    R_{SG}\left( x \right) = \underset{x}{\arg\min} || C_{SG}(F^{-1}y;\hat{\theta}_{SG}) - x||^2_2
    \label{eq:m1hatclosed}
\end{align}
where $C_{SG}$ is an image-domain deep network with learned parameters $\hat{\theta}_{SG}$. The formulation in Eq. \ref{eq:m1hatclosed} assumes that $C_{SG}$ recovers multi-coil output images provided multi-coil input images. The parameters ${\theta}_{SG}$ for $C_{SG}$ can be learned based on a pixel-wise loss between reconstructed and ground-truth images. Training is conducted offline via an empirical risk minimization approach based on Monte Carlo sampling \cite{Schlemper2017}:
\begin{align}
    \mathcal{L}_{SG}(\theta_{SG}) =  \sum_{n=1}^{N} || C_{SG}(F^{-1}y^n;\theta_{SG}) - \breve{x}^{n}||_p 
    \label{eq:lossrec}
\end{align}
where $N$ is the number of training scans, $n$ is the training scan index, $||.||_p$  denotes $\ell_p$ norm, $\breve{x}^{n}$ is the ground-truth multi-coil image derived from the fully-sampled acquisition for the $n$th scan, and $y^{n}$ are respective undersampled k-space data.

A common approach to build $C_{SG}$ is based on unrolled architectures that perform cascaded projections through CNN blocks to regularize the image and DC blocks to ensure conformance to the physical signal model \cite{MoDl}. Given a total of $K$ cascades with tied CNN parameters across cascades, the mapping through the $k$th cascade is \cite{Schlemper2017,huang2021dynamic,huang2022evaluation}:
\begin{align}
    x^r_{k}=f_{DC}\left( f_{SG}\left( x^r_{k-1};{\theta}_{SG} \right) \right)
    \label{eq:fdcserialcsg}
\end{align}
where $x_k^r$ is the image for the $r$th scan (that could be a training or test scan) at the output of the $k$th cascade ($k \in [1,2,...,K]$), and $x_{0}^r = F^{-1}y^r$ where $y^r$ are the acquired undersampled data for the $r$th scan. Meanwhile, $f_{SG}$ is the CNN block embedded in the $k$th cascade with parameters ${\theta}_{SG}$. 

As the parameters of SG priors are trained offline and then frozen during inference, deeper network architectures can be used for enhanced reconstruction performance along with fast inference. However, learning deep networks requires substantial training datasets that may be difficult to collect. Moreover, since SG priors learn aggregate representations of MRI data across training subjects, they may show poor generalization to subject-specific variability in anatomy \cite{Yilmaz2021}.

\textbf{\textit{SS priors}}: Unlike SG priors, SS priors are not learned from a dedicated training dataset but instead they are learned directly for individual test scans to improve generalization \cite{raki}. The SS prior can also be used to implement $R(.)$ in Eq. \ref{eq:rec_reg_1} with the respective regularization term expressed as:
\begin{align}
    R_{SS}\left( {x}  \right)= \underset{x}{\arg\min} ||C_{SS}(F^{-1}y;\hat{\theta}_{SS}) - x||^2_2
    \label{eq:m2hatclosed}
\end{align}
where $C_{SS}$ is an image-domain network with parameters $\hat{\theta}_{SS}$. In the absence of ground-truth images, the parameters ${\theta}_{SS}^q$ for the $q$th test scan can be learned based on proxy k-space losses between reconstructed and acquired undersampled data \cite{Knoll2019inverseGANs}. Learning is conducted online to minimize this proxy loss: 
\begin{align}
    \mathcal{L}_{SS}(\theta_{SS}^q) = || MF C_{SS}(F^{-1}y^{q};\theta_{SS}^q) - y^q||_p
    \label{eq:raki}
\end{align}
where $y^{q}$ are acquired undersampled k-space data for the $q$th scan. An unrolled architecture can be adopted to build $C_{SS}$ by performing cascaded projections through network and DC blocks, resulting in the following mapping for the $k$th cascade:
\begin{align}
    x^q_{k}=f_{DC}\left( f_{SS}\left( x^q_{k-1};\theta_{SS}^q \right) \right)
    \label{eq:fdcserialcsg}
\end{align}
$f_{SS}$ can be operationalized as a linear or nonlinear network \cite{Konukoglu2019,Knoll2019inverseGANs}. As the parameters of SS priors are learned independently for each test scan, they promise enhanced generalization to subject-specific anatomy. However, since training is performed online during inference, SS priors can introduce substantial computational burden, particularly when deep nonlinear networks are used that also increase the risk of overfitting \cite{DIP}. 

\begin{figure*}[h!]   
  \centering
\includegraphics[width=0.75\linewidth]{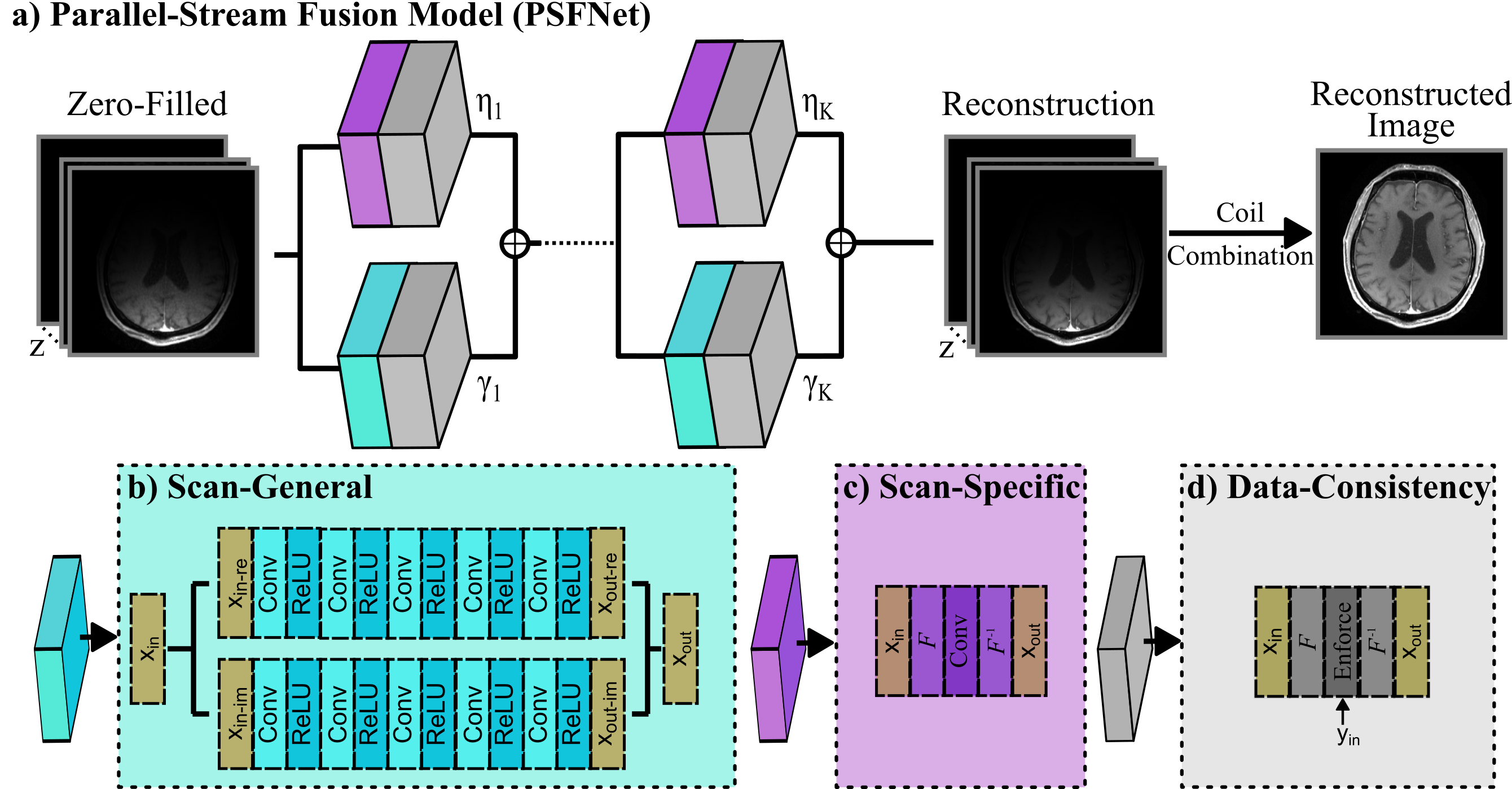}  
\caption{\textbf{(a)} PSFNet comprises a parallel-stream cascade of sub-networks where each sub-network contains \textbf{(b)} a scan-general (SG) block, and \textbf{(c)} a scan-specific (SS) block. The two parallel blocks are each succeeded by \textbf{(d)} a data-consistency (DC) block, and their outputs are aggregated with learnable fusion weights, $\eta_k$ and $\gamma_k$ where $k$ is the cascade index. At the end of $K$ cascades, coil-combination is performed on multi-coil data using sensitivity maps estimated via ESPIRiT \cite{Uecker2014}. The SG block is implemented as a deep convolutional neural network (CNN) and the SS block was implemented as a linear projection layer.}
  \label{fig:main_fig}
\end{figure*}

\subsection{PSFNet}
Here, we propose to combine SS and SG priors to maintain a favorable trade-off between generalization performance and computational efficiency under low-data regimes. In the conventional unrolling framework, this requires computation of serially alternated projections through the SS, SG and DC blocks:
\begin{align}
   x^r_{k}=f_{DC}\left( f_{SG}\left( f_{SS}\left(x^r_{k-1};\theta^r_{SS} \right);{\theta}_{SG} \right) \right)
    \label{eq:fdcserialcsg2}
\end{align}
The unrolled architecture with $K$ cascades can be learned offline using the training set. Note that scarcely-trained SG blocks under low-data regimes can perform suboptimally, introducing residual errors in their output. In turn, these errors will accumulate across serial projections to degrade the overall performance.  

To address this limitation, here we introduce a novel architecture, PSFNet, that performs parallel-stream fusion of SS and SG priors as opposed to the serial combination in conventional unrolled methods. PSFNet utilizes a nonlinear SG prior for high performance, and a linear SS prior to enhance generalization without excessive computational burden. The two priors undergo parallel-stream fusion with learnable fusion parameters $\eta$ and $\gamma$, as displayed in Figure \ref{fig:main_fig}. These parameters adaptively control the relative weighting of information extracted by the SG versus SS streams during the course of training in order to alleviate error accumulation. As such, the mapping through the $k$th cascade in PSFNet is:
\begin{multline}
x^r_k=   \eta_k f_{DC}( f_{SS} (x^r_{k-1};\theta_{SS}^r)) + \gamma_k f_{DC}(f_{SG}(x^r_{k-1};{\theta}_{SG}))
\label{eq:hybrid_forward}
\end{multline}
In Eq. \ref{eq:hybrid_forward}, the learnable fusion parameters for the SS and SG blocks at the $k$th cascade are $\eta_k$ and $\gamma_k$, respectively. To enforce fidelity to acquired data, DC projections are performed on the outputs of SG and SS blocks. In PSFNet, the SG prior is learned collectively from the set of training scans and then frozen during inference on test scans. In contrast, the SS prior is learned individually for each scan, during both training and inference.

\underline{\textit{Training}}:
PSFNet involves a training phase to learn model parameters for the SG prior as well as its fusion with the SS prior. For each individual scan in the training set, PSFNet learns a dedicated SS prior for the given scan. Since learning of a nonlinear SS prior has substantial computational burden, we adopt a linear SS prior in PSFNet. In particular, the SS block performs dealiasing via convolution with a linear kernel \cite{Uecker2014}:
\begin{align}
    f_{SS}(x^n_{k-1};\theta^n_{SS})= F^{-1} \{ \theta^n_{SS} \circledast F x^n_{k-1} \}
    \label{eq:linearss}
\end{align}
where $\theta^n_{SS} \in \mathbb{C}^{(z \times z \times w \times w)}$ with $n$ denoting the training scan index, $z$ denoting the number of coil elements, and $w$ denoting the kernel size in k-space. The SS blocks contain unlearned Fourier and inverse Fourier transformation layers as their input and output layers, respectively, and convolution is computed over the spatial frequency dimensions in k-space. Meanwhile, the SG prior is implemented as a deep CNN operating in image domain:
\begin{align}
    f_{SG}(x^n_{k-1};\theta_{SG})= CNN(x^n_{k-1})
    \label{eq:nonlinearsg}
\end{align}
Across the scans in the training set, the training loss for PSFNet can then be expressed in constrained form as:
\begin{multline}
    \mathcal{L}_{PSFNet}(\theta_{SG},\pmb{\gamma},\pmb{\eta}) =
    \sum_{n=1}^{N}  || \eta_{K}  f_{DC} (f_{SS}(x^n_{K-1};\hat{\theta}_{SS}^n)) \\+
     \gamma_{K} f_{DC}(f_{SG}(x^n_{K-1};\theta_{SG})) - \breve{x}^n||_p  
    \label{eq:lossrec} \\
    \mbox{s.t. }  \hat{\theta}_{SS}^n = \underset{\theta_{SS}^n}{\arg\min} || F^{-1}W^ny^n-f_{SS}(F^{-1}W^n y^n;\theta^n_{SS}) ||_2^2
\end{multline}
The constraint in Eq. \ref{eq:lossrec} corresponds to the scan-specific learning of the SS prior $\hat{\theta}^n_{SS}$, which is then adopted to calculate the loss. Assuming that the linear relationships among neighboring spatial frequencies are similarly distributed across k-space \cite{Uecker2014}, $\hat{\theta}^n_{SS}$ is learned by solving a self-regression problem on the subset of fully-sampled data in central k-space,
where $W^n$ is a mask operator that selects data within this calibration region. 

Note that, unlike deep reconstruction models purely based on SG priors, the SG prior in PSFNet is not directly trained to remove artifacts in zero-filled reconstructions of undersampled data. Instead, the SG prior is trained to concurrently suppress artifacts in reconstructed images along with the SS prior; and the relative importance attributed to the two priors is determined by the fusion parameters at each cascade. As such, the SS prior can be given higher weight during initial training iterations where the SG prior is scarcely trained, whereas its weight can be relatively reduced during later iterations once the SG prior has been sufficiently trained. This adaptive fusion approach thereby lowers reliance on the availability of large training sets.


\underline{\textit{Inference}}: During inference on the $q$th test scan, the respective SS prior is learned online as:
\begin{align}
\hat{\theta}_{SS}^q = \underset{\theta_{SS}^q}{\arg\min} || F^{-1}W^qy^q-f_{SS}^q(F^{-1}W^qy^q;\theta^q_{SS}) ||_2^2
    \label{eq:spirit_1_2}
\end{align}
Afterwards, the learned $\hat{\theta}_{SS}^q$ is used along with the previously trained $\hat{\theta}_{SG}$ to perform repeated projections through $K$ cascades as described in Eq. \ref{eq:hybrid_forward}. The multi-coil image recovered by PSFNet at the output of the $K$ cascade is:
\begin{multline}
\hat{x}^q=  \eta_K f_{DC}(f_{SS} (x^q_{K-1};\hat{\theta}_{SS}^q)) + \gamma_K  f_{DC}(f_{SG}(x_{K-1}^q;\hat{\theta}_{SG}))
    \label{eq:fdcserialcsg23}
\end{multline}
where $\hat{x}^q$ denotes the recovered image. The final reconstruction can be obtained by performing combination across coils:
\begin{equation}
\hat{x}^q_{combined} = A^* \hat{x}^q  
\end{equation}
where $A$ are coil sensitivities, and $A^*$ denotes the conjugate of $A$.

\section{Methods}
\subsection{Implementation Details}
In each cascade, PSFNet contained two parallel streams with SG and SS blocks. The SG blocks comprised an input layer followed by a stack of 4 convolutional layers with 64 channels and 3x3 kernel size each, and an output layer with ReLU activation functions. They processed complex images with separate channels for real and imaginary components. The SS blocks comprised a Fourier layer, 5 projection layers with identity activation functions, and an inverse Fourier layer. They processed complex images directly without splitting real and imaginary components. The linear convolution kernel used in the projection layers was learned from the calibration region by solving a Tikhonov regularized self-regression problem \cite{Lustig2010}. The DC blocks comprised 3 layers respectively to implement forward Fourier transformation, restoration of acquired k-space data and inverse Fourier transformation. PSFNet was implemented with 5 cascades, $K$=5. The weights of SG, SS, and DC blocks were tied across cascades to limit model complexity \cite{MoDl}. The only exception were fusion coefficients that determine the relative weighting of the SG and SS blocks at each stage ($\gamma_1,..,\gamma_k,...,\gamma_5$ $\eta_1,...\eta_k,...,\eta_5$). These fusion parameters were kept distinct across cascades. Coil-combination on the recovered multi-coil images was performed using sensitivity maps estimated via ESPIRiT \cite{Uecker2014}. 

\subsection{MRI Dataset}
Experimental demonstrations were performed using brain MRI scans from the NYU fastMRI database \cite{fastmri}. Here, contrast-enhanced T\SB{1}-weighted (cT\SB{1}-weighted) and T\SB{2}-weighted acquisitions were considered. The fastMRI dataset contains volumetric MRI data with varying image and coil dimensionality across subjects. Note that a central aim of this work was to systematically examine the learning capabilities of models for varying number of training samples. To minimize potential biases due to across-subject variability in MRI protocols, here we selected subjects with matching imaging matrix size and number of coils. To do this, we only selected subjects with at least 10 cross-sections and 
only the central 10 cross-sections were retained in each subject. We further selected subjects with an in-plane matrix size of 256x320 for cT\SB{1} acquisitions, and of 288x384 for T\SB{2} acquisitions. Background regions in MRI data with higher dimensions were cropped. Lastly, we restricted our sample selection to subjects with at least 5 coil elements, and geometric coil compression \cite{Zhang2013} was applied to unify the number of coils to 5 in all subjects.

Fully-sampled acquisitions were retrospectively undersampled to achieve acceleration rates of R=4x and 8x. Random undersampling patterns were designed via either a bi-variate normal density function peaking at the center of k-space, or a uniform density function across k-space. The standard deviation of the normal density function was adjusted to maintain the expected value of R across k-space. The fully-sampled calibration region spanned a 40x40 window in central k-space.

\subsection{Competing Methods}
PSFNet was compared against several state-of-the-art approaches including SG methods, SS methods, and traditional PI reconstructions. For methods containing SG priors, both supervised and unsupervised variants were implemented. 

\textbf{PSFNet}: A supervised variant of PSFNet was trained using paired sets of undersampled and fully-sampled acquisitions. 

\textbf{PSFNet\SB{US}}: An unsupervised variant of PSFNet was implemented using self-supervision based on only undersampled training data. Acquired data were split into two non-overlapping sets where 40\% of samples was reserved for evaluating the training loss and 60\% of samples was reserved to enforce DC \cite{yaman2020}.

\textbf{MoDL}: A supervised SG methods based on an unrolled architecture with tied weights across cascades was used \cite{MoDl}. MoDL serially interleaves SG and DC blocks. The number of cascades and the structure of SG and DC blocks were identical to those in PSFNet. 

\textbf{MoDL\SB{US}}: An unsupervised variant of MoDL was implemented using self-supervision. A 40\%-60\% split was performed on acquired data to evaluate the training loss and enforce data consistency, respectively \cite{yaman2020}.

\textbf{sRAKI-RNN}: An SS method was implemented based on the MoDL architecture \cite{sraki_rnn}. Learning was performed to minimize DC loss on the fully-sampled calibration region. Calibration data were randomly split with 75\% of samples used to define the training loss and 25\% of samples reserved to enforce DC. Multiple input-output pairs were produced for a single test sample by utilizing this split.

\textbf{SPIRiT}: A traditional PI reconstruction was performed using the SPIRiT method \cite{Lustig2010}. Reconstruction parameters including the regularization weight for kernel estimation ($\kappa$), kernel size ($w$), and the number of iterations ($N_{iter}$) were independently optimized for each reconstruction task via cross-validation. 

\textbf{SPARK}: An SS method was used to correct residual errors from an initial SPIRiT reconstruction \cite{spark}. Learning was performed to minimize DC loss on the calibration region. The learned SS prior was then used to correct residual errors in the remainder of k-space.

\subsection{Optimization Procedures} 
For all methods, hyperparameter selection was performed via cross-validation on a three-way split of data across subjects. There was no overlap among training, validation and test sets in terms of subjects. Data from 10 subjects were reserved for validation, and data from a separate set of 40 subjects were reserved for testing. The number of subjects in the training set was varied from 1 to 50. Hyperparameters that maximized peak signal-to-noise ratio (PSNR) on the validation set were selected for each method. 

Training was performed via the Adam optimizer with learning rate $\zeta$=$10^{-4}$, $\beta_1$=0.90 and $\beta_2$=0.99 \cite{Kingma2015}. All deep learning methods were trained to minimize hybrid $\ell_1$-$\ell_2$-norm loss between recovered and target data (e.g., between reconstructed and ground truth images for PSFNet, between recovered and acquired k-space samples for PSFNet\SB{US}) \cite{yaman2020}. For PSFNet and MoDL, the selected number of epochs was 200, batch size was set to 2 for the limited number of training samples ($N_{samples}<$10), and to 5 otherwise. In DC blocks, $\lambda=\infty$ was used to enforce strict data consistency. 
For PSFNet and SPIRiT, the kernel width ($w$) and regularization parameter ($\kappa$) values were set as ($\kappa$, $w$) = ($10^{-2}$, 9) at R= 4 and ($10^{-2}$, 9) at R=8 for cT\SB{1}-weighted reconstructions, and as (10\SP{0}, 17) at R=4 and ($10^{-2}$, 17) at R=8 for T\SB{2}-weighted reconstructions. For SPIRiT, the number of iterations $N_{iter}$ was set as 13 at R=4 and 27 at R=8 for cT\SB{1}-weighted reconstructions, 20 at R=4 and 38 at R=8 for T\SB{2}-weighted reconstructions. 
For sRAKI-RNN, the selected number of epochs was 500 and batch size was set to 32. All other optimization procedures were identical to MoDL. 
For SPARK, network architecture and training procedures were adopted from \cite{spark}, except for the number of epochs ($N_{epoch}$) and learning rate ($\zeta$) which were optimized on the validation set as ($N_{epoch}$, $\zeta$)= (100, $10^{-2}$) For cT\SB{1}-weighted reconstructions, and ($N_{epoch}$, $\zeta$)= (250, $10^{-3}$) for T\SB{2}-weighted reconstructions.

All competing methods were executed on an NVidia RTX 3090 GPU, and models were coded in Tensorflow except for SPARK which was implemented in PyTorch. SPARK was implemented using the toolbox at \url{https://github.com/YaminArefeen/spark_mrm_2021}. The code to implement PSFNet will be available publicly at \url{https://github.com/icon-lab/PSFNet} upon publication.

\subsection{Performance Metrics} 
Performance assessments for reconstruction methods were carried out by visual observations and quantitative metrics. PSNR and structural similarity index (SSIM) were used for quantitative evaluation. For each method, metrics were computed on coil-combined images from the reconstruction and from the fully-sampled ground truth acquisition. Statistical differences between competing methods were examined via non-parametric Wilcoxon signed-rank tests.

\begin{figure}[t]
\centering
\includegraphics[width=0.9\linewidth]{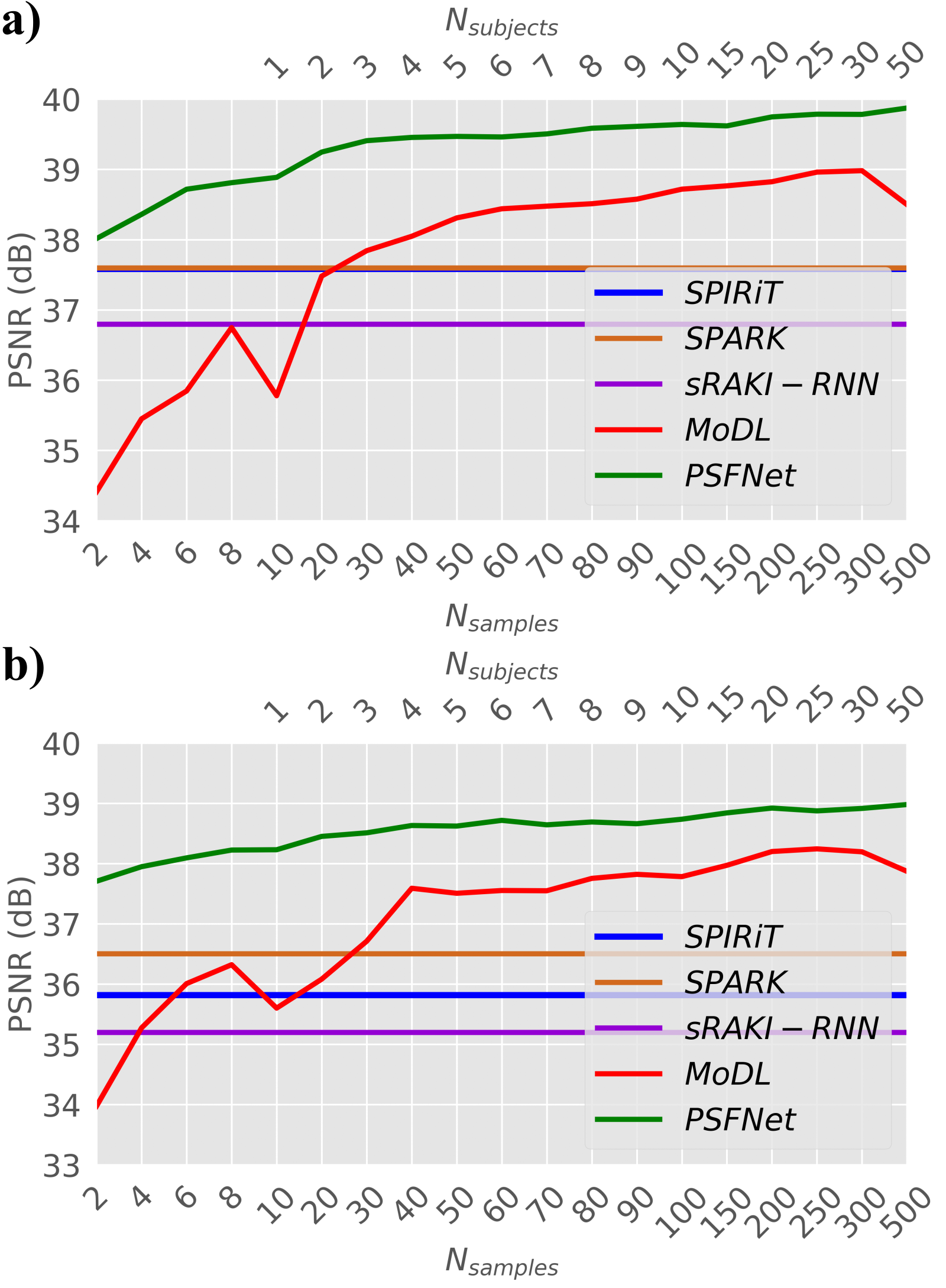}
\caption{Average PSNR across test subjects for \textbf{(a)} cT\textsubscript{1}- and \textbf{(b)} T\textsubscript{2}-weighted image reconstructions at R=4x. Model training was performed for varying number of training samples ($N_{samples}$, lower x-axis) and thereby training subjects ($N_{subjects}$, upper x-axis). Results are shown for SPIRiT, SPARK, sRAKI-RNN, MoDL and PSFNet.}
\label{fig:plot_subs}
\end{figure}

\begin{figure*}[t]
\centering
\includegraphics[width=0.7\linewidth]{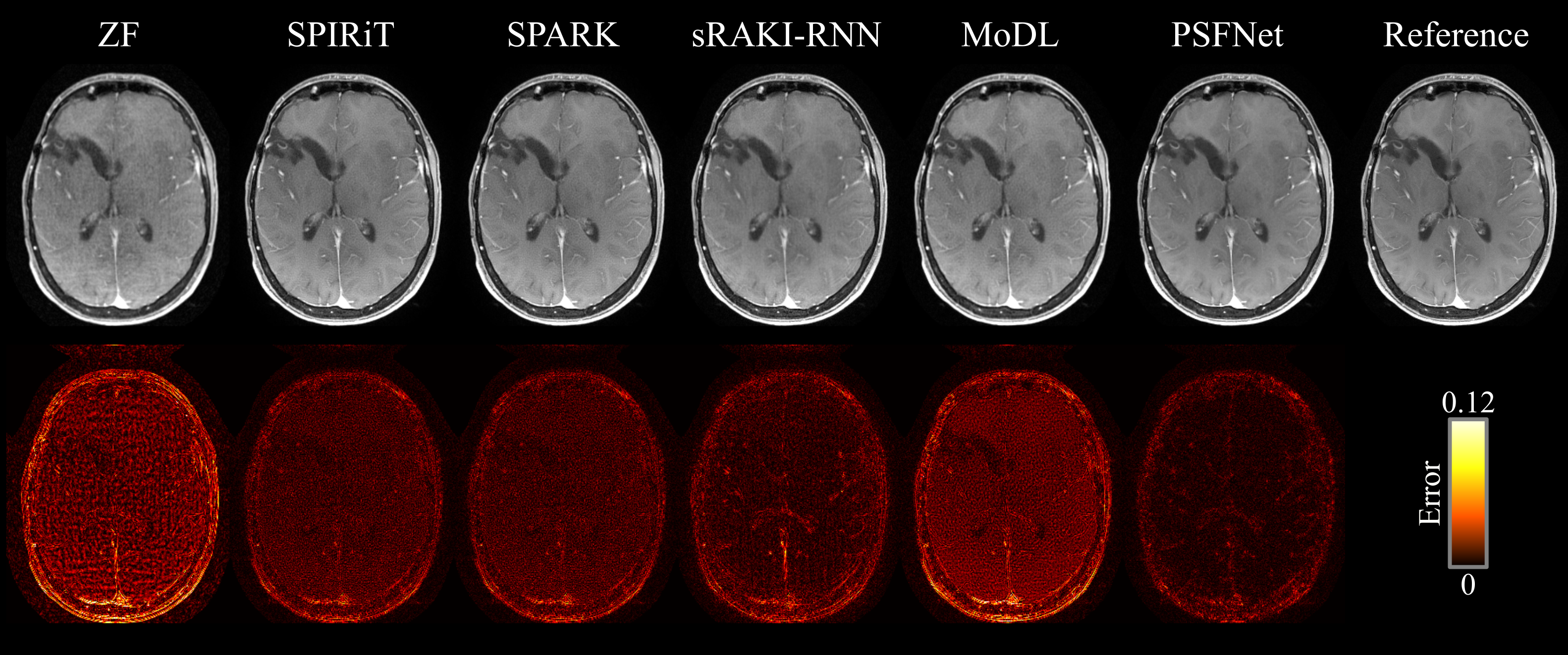}
\caption{cT\SB{1}-weighted image reconstructions at R=4x via SPIRiT, SPARK, sRAKI-RNN, MoDL, and PSFNet along with the zero-filled reconstruction (ZF) and the reference image obtained from the fully-sampled acquisition. Error maps for each method are shown in the bottom row. MoDL and PSFNet were trained on 10 cross-sections from a single subject. SPIRiT, SPARK and sRAKI-RNN directly performed inference on test data without a priori model training. PSFNet shows superior performance to competing methods in terms of residual reconstruction errors.}
\label{fig:T1c}
\end{figure*}

\begin{figure*}[t]
\centering
\includegraphics[width=0.7\linewidth]{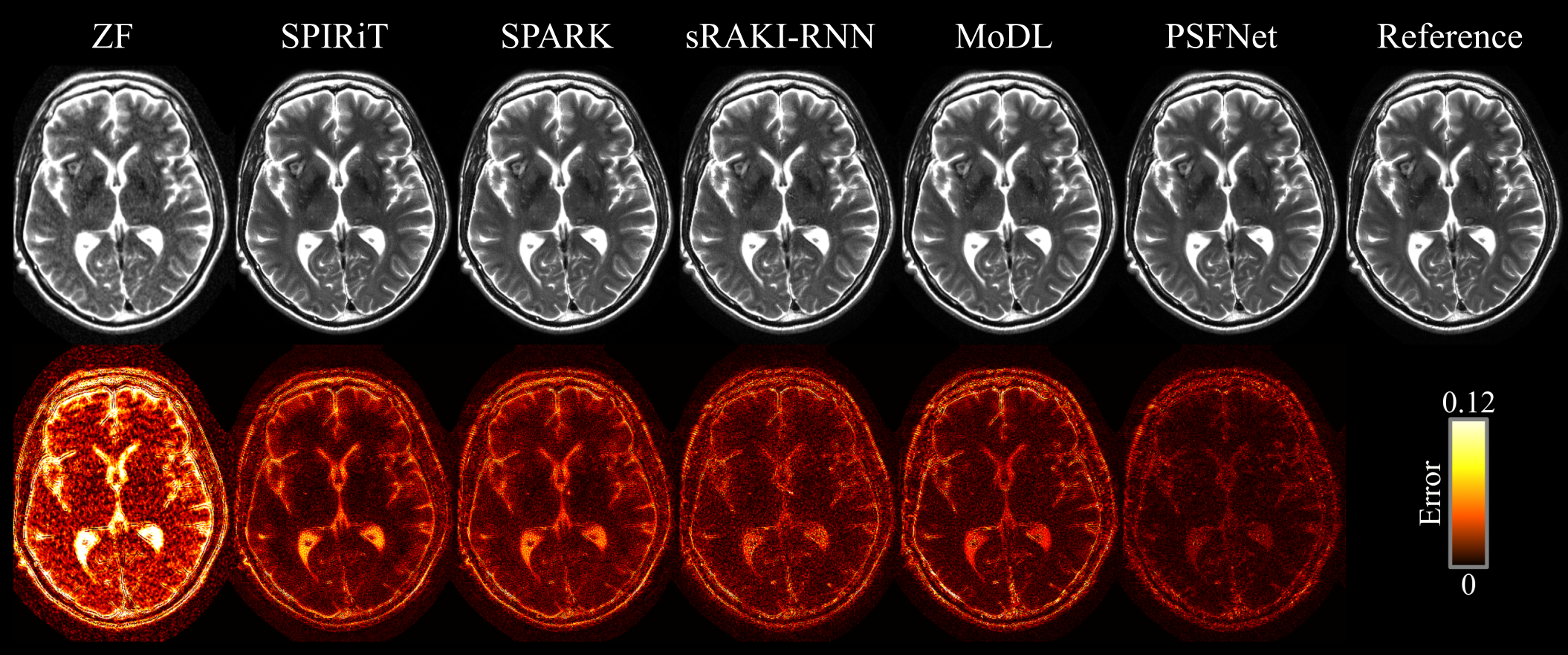}
\caption{T\SB{2}-weighted image reconstructions at R=4x via SPIRiT, SPARK, sRAKI-RNN, MoDL, and PSFNet along with the zero-filled reconstruction (ZF) and the reference image obtained from the fully-sampled acquisition. Error maps for each method are shown in the bottom row. MoDL and PSFNet were trained on 10 cross-sections from a single subject. SPIRiT, SPARK and sRAKI-RNN directly performed inference on test data without a priori model training. PSFNet shows superior performance to competing methods in terms of residual reconstruction errors.}
\label{fig:T2}
\end{figure*}

\subsection{Experiments}
Several different experiments were conducted to systematically examine the performance of competing methods. Assessments aimed to investigate reconstruction performance under low training data regimes, generalization performance in case of mismatch between training and testing domains, contribution of the parallel-stream design to reconstruction performance, sensitivity to hyperparameter selection, performance in unsupervised learning, and computational complexity. 

\textbf{Performance in low-data regimes}: Deep SG methods for MRI reconstruction typically suffer from suboptimal performance as the size of the training dataset is constrained. To systematically examine reconstruction performance, we trained supervised variants of PSFNet and MoDL while the number of training samples ($N_{samples}$) was varied in the range [2-500] cross sections. To attain a given number of samples, sequential selection was performed across subjects and across cross-sections within each subject. Thus, the number of unique subjects included in the training set roughly corresponded to $N_{samples}/10$ (since there were 10 cross-sections per subject). SS reconstructions were also performed with sRAKI-RNN, SPIRiT and SPARK. In the absence of fully-sampled ground truth data to guide the learning of the prior, unsupervised training of deep reconstruction models may prove relatively more difficult compared to supervised training. In turn, this may elevate requirements on training datasets for unsupervised models. To examine data efficiency for unsupervised training, we compared the reconstruction performance of PSFNet\SB{US} and MoDL\SB{US} as $N_{samples}$ was varied in the range of [2-500] cross sections. Comparisons were also provided against sRAKI-RNN, SPIRiT and SPARK. 

\textbf{Generalization performance}: Deep reconstruction models can suffer from suboptimal generalization when the MRI data distribution shows substantial variation between the training and testing domains. To examine generalizability, PSFNet models were trained on data from a source domain and tested on data from a different target domain. The domain-transferred models were then compared to models trained and tested directly in the target domain. Three different factors were altered to induce domain variation: tissue contrast, undersampling pattern, and acceleration rate. First, the capability to generalize to different tissue contrasts was evaluated. Models were trained on data from a source contrast and tested on data from a different target contrast. Domain-transferred models were compared to target-domain models trained on data from the target contrast. Next, the capability to generalize to different undersampling patterns was assessed. Models were trained on data undersampled with variable-density patterns and tested on data undersampled with uniform-density patterns. Domain-transferred models were compared to target-domain models trained on uniformly undersampled data. Lastly, the capability to generalize to different acceleration rates was examined. Models were trained on acquisitions accelerated at R=4x and tested on acquisitions accelerated at R=8x. Domain-transferred models were compared to target-domain models trained at R=8x. 

\textbf{Sensitivity to hyperparameters}:
SS priors are learned from individual test scans as opposed to SG priors trained on larger training datasets. Thus, SS priors might show elevated sensitivity to hyperparameter selection. We assessed the reliability of reconstruction performance against suboptimal hyperparameter selection for SS priors. For this purpose, analyses were conducted on SPIRiT, SPARK and PSFNet that embody SS methods to perform linear reconstructions in k-space. The set of hyperparameters examined included regularization parameters for kernel estimation ($\kappa$) and kernel size ($w$). Separate models were trained using $\kappa$ in range [10\textsuperscript{-3}-10\textsuperscript{0}] and $w$ in range [5-17]. 

\textbf{Computational complexity}:
Finally, we assessed the computational complexity of competing methods. For each method, training and inference times were measured for a single subject with 10 cross-sections. Each cross-section had an imaging matrix size of 256x320 and contained data from 5 coils. For all methods including SS priors, hyperparameters optimized for cT\SB{1}-weighted reconstructions at R=4 were used.

\textbf{Ablation analysis}:
To assess the contribution of the parallel-stream design in PSFNet, a conventional unrolled variant of PSFNet was formed, named as PSFNet\SB{Serial}. PSFNet\SB{Serial} combined the SG and SS priors via serial projections as described in Eq. \ref{eq:fdcserialcsg2}. Modeling procedures and the design of SG and SS blocks were kept identical between PSFNet and PSFNet\SB{Serial} for fair comparison. Performance was assessed as $N_{samples}$ was varied in the range of [2-500] cross sections.

\section{Results}

\subsection{Performance in Low-Data Regimes}
Common SG methods for MRI reconstruction are based on deep networks that require copious amounts of training data, so performance can substantially decline on limited training sets \cite{Dar2017,KnollGeneralization}. In contrast, PSFNet leverages an SG prior to concurrently reconstruct an image along with an SS prior. Therefore, we reasoned that its performance should scale favorably under low-data regimes compared to SG methods. We also reasoned that PSFNet should yield elevated performance compared to SS methods due to residual corrections from its SG prior. To test these predictions, we trained supervised variants of PSFNet and MoDL along with SPIRiT, sRAKI-RNN, and SPARK while the number of training samples ($N_{samples}$) was systematically varied. Figure \ref{fig:plot_subs} displays PSNR performance for cT\textsubscript{1}-weighted and T\textsubscript{2}-weighted image reconstruction as a function of $N_{samples}$. PSFNet outperforms the scan-general MoDL method for all values of $N_{samples}$ ($p<0.05$). As expected, performance benefits with PSFNet become more prominent towards lower values of $N_{samples}$. PSFNet also outperforms traditional SPIRiT and scan-specific sRAKI-RNN and SPARK methods broadly across the examined range of $N_{samples}$ ($p<0.05$). Note that while MoDL requires $N_{samples}=30$ (3 subjects) to offer on par performance to SS methods, PSFNet yields superior performance with as few as $N_{samples}=2$. Representative reconstructions for cT\textsubscript{1}- and T\textsubscript{2}-weighted images are depicted in Figures \ref{fig:T1c} and \ref{fig:T2}, where $N_{samples}=10$ from a single subject were used for training. PSFNet yields lower reconstruction errors compared to all other methods in this low-data regime, where competing methods either show elevated noise or blurring.

\begin{figure}[t]
\centering
\includegraphics[width=\linewidth]{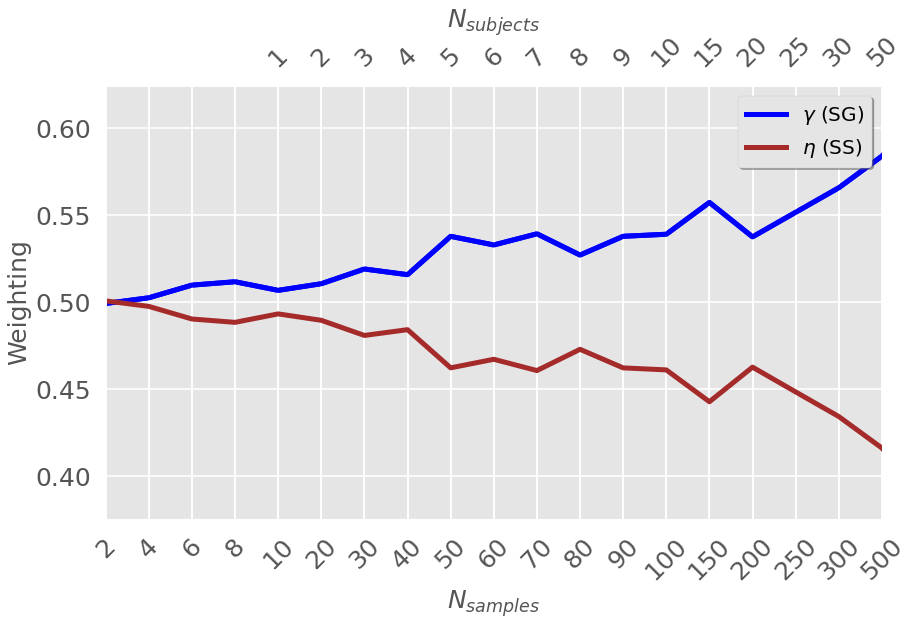}
\caption{Weighting of the SG ($\gamma$) and SS ($\eta$) blocks in the final cascade of PSFNet. Weights were averaged across models trained for  cT\textsubscript{1}- and T\textsubscript{2}-weighted reconstructions at R=4x. Model training was performed for varying number of training samples ($N_{samples}$, lower x-axis) and thereby training subjects ($N_{subjects}$, upper x-axis). Both blocks are equally weighted with very limited training data. As $N_{samples}$ increases, the weighting of the SG prior becomes more dominant over the weighting of the SS prior.}
\label{fig:ss_sg_weightings}
\end{figure}

\begin{figure}[t]
\includegraphics[width=0.9\linewidth]{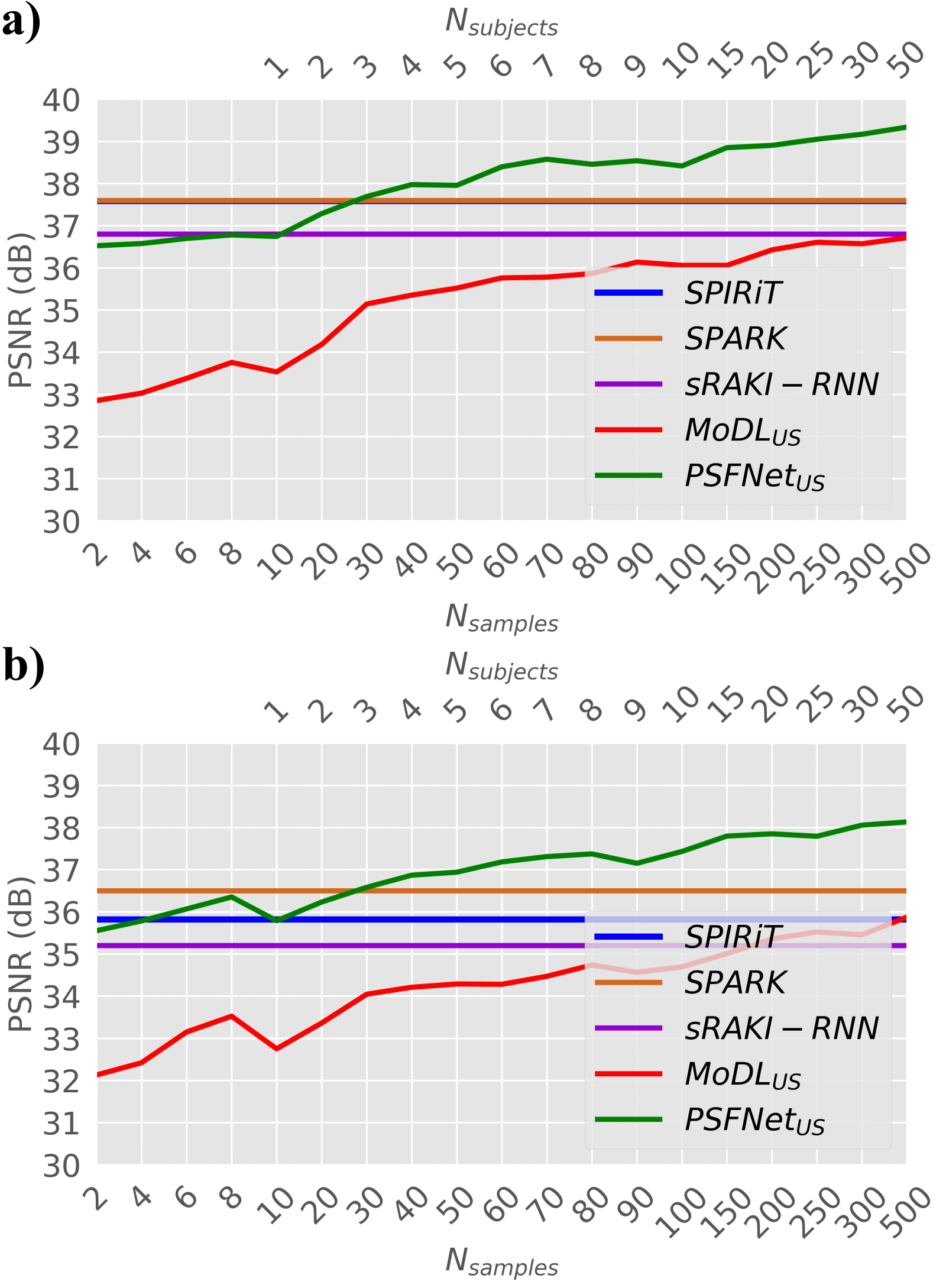}
\caption{Average PSNR across test subjects for \textbf{(a)} cT\textsubscript{1}- and \textbf{(b)} T\textsubscript{2}-weighted image reconstructions at R=4x. Model training was performed for varying number of training samples ($N_{samples}$, lower x-axis) and thereby training subjects ($N_{subjects}$, upper x-axis). Results are shown for SPIRiT, SPARK, sRAKI-RNN, MoDL\SB{US} and PSFNet\SB{US}.}
\label{S-fig:plot_subs_us}
\end{figure}

Naturally, the performance of PSFNet increases as more training samples are available. Since the SS prior is independently learned for individual samples, it should not elicit systematic performance variations depending on $N_{samples}$. Thus, the performance gains can be attributed to improved learning of the SG prior. In turn, we predicted that PSFNet would put more emphasis on its SG stream as its reliability increases. To examine this issue, we inspected the weightings of the SG ($\gamma$) and SS ($\eta$) streams as the training set size was varied. Figure \ref{fig:ss_sg_weightings} displays weightings at the last cascade as a function of $N_{samples}$. For lower values of $N_{samples}$ where the quality of the SG prior is relatively limited, the SG and SS priors are almost equally weighted. In contrast, as the learning of the SG prior improves with higher $N_{samples}$, the emphasis on the SG prior increases while the SS prior is less heavily weighted.

\begin{figure*}[t] 
\centering
\includegraphics[width=0.7\linewidth]{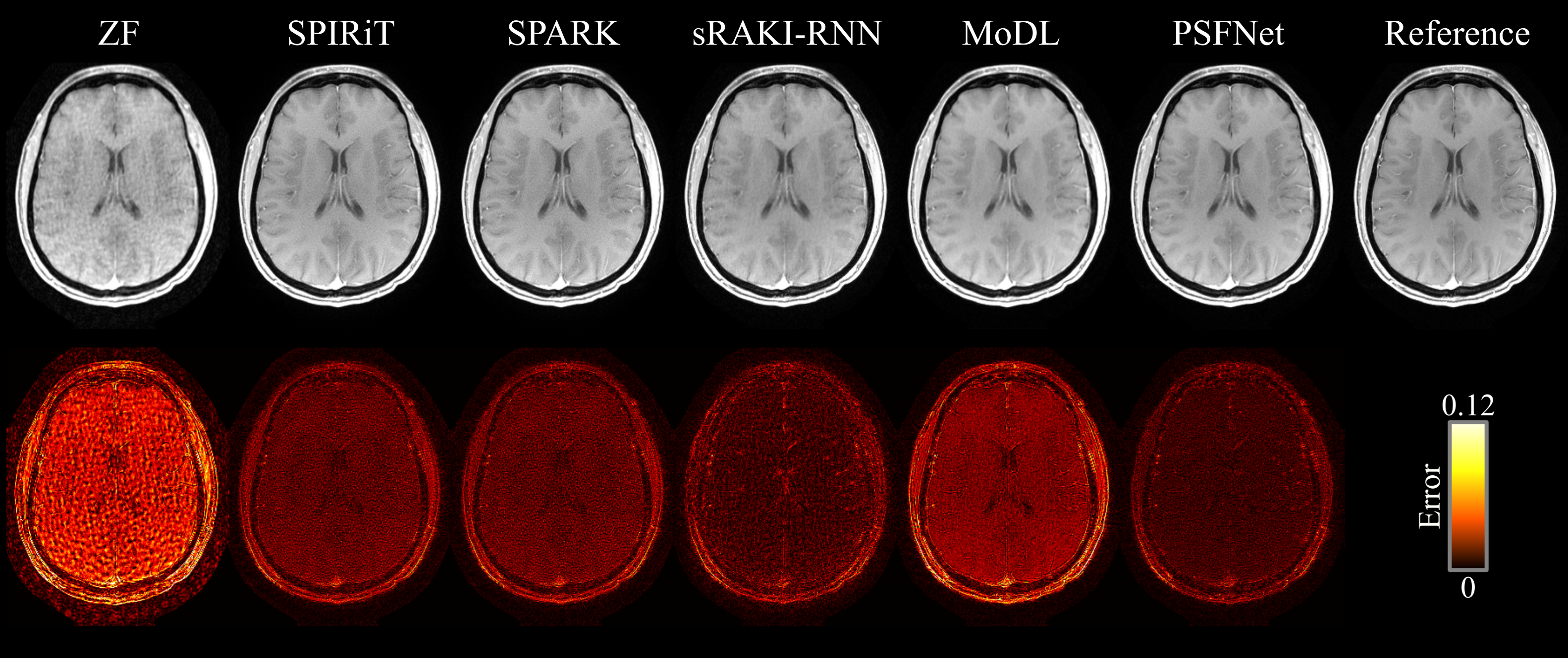}
\caption{cT\SB{1}-weighted image reconstructions at R=4x via SPIRiT, SPARK, sRAKI-RNN, MoDL\SB{US}, and PSFNet\SB{US} along with the zero-filled reconstruction (ZF) and the reference image obtained from the fully-sampled acquisition. Error maps for each method are shown in the bottom row. MoDL\SB{US} and PSFNet\SB{US} were trained on 100 cross-sections (from 10 subjects). SPIRiT, SPARK and sRAKI-RNN directly performed inference on test data without a priori model training. PSFNet\SB{US} shows superior performance to competing methods in terms of residual reconstruction errors. }
\label{S-fig:T1c_us}
\end{figure*}

\begin{figure*}[t]
\centering
\includegraphics[width=0.7\linewidth]{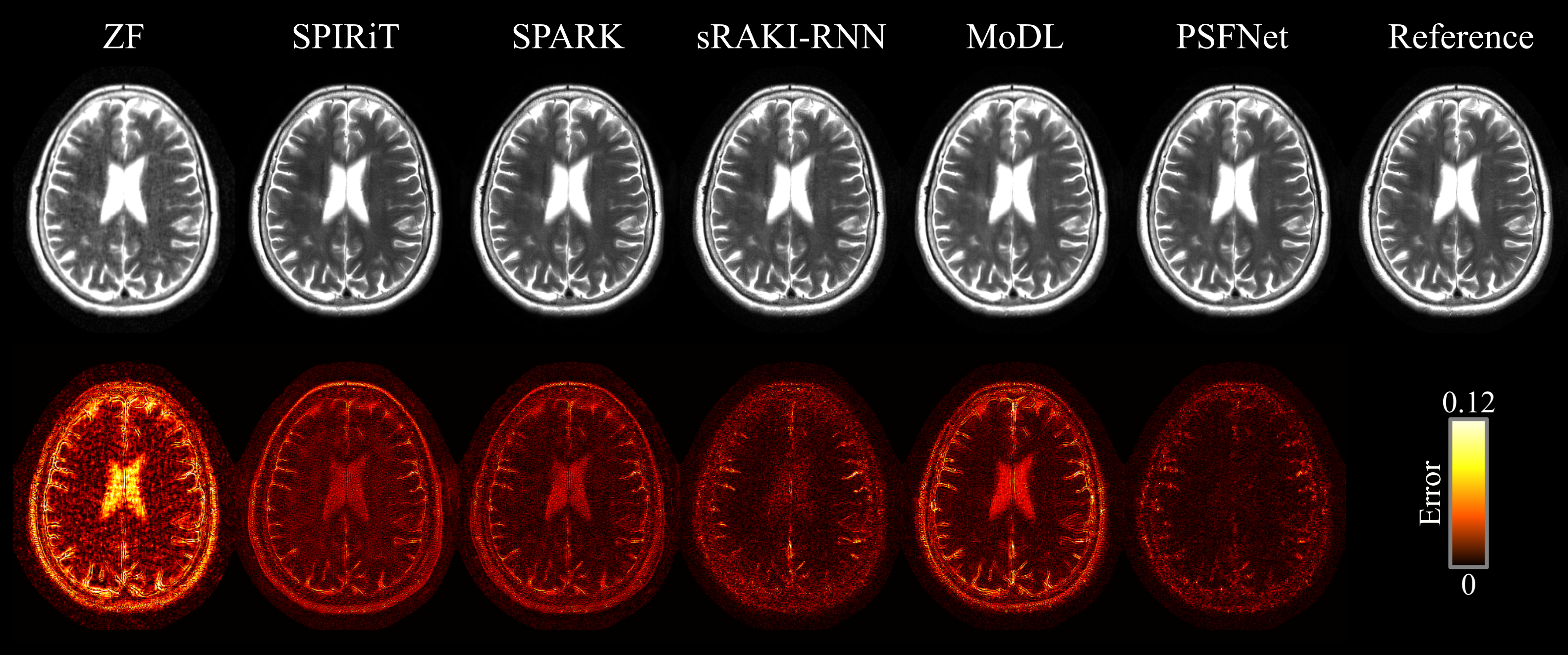}
\caption{T\SB{2}-weighted image reconstructions at R=4x via SPIRiT, SPARK, sRAKI-RNN, MoDL\SB{US}, and PSFNet\SB{US} along with the zero-filled reconstruction (ZF) and the reference image obtained from the fully-sampled acquisition. Error maps for each method are shown in the bottom row. MoDL\SB{US} and PSFNet\SB{US} were trained on 100 cross-sections (from 10 subjects). SPIRiT, SPARK and sRAKI-RNN directly performed inference on test data without a priori model training. PSFNet\SB{US} shows superior performance to competing methods in terms of residual reconstruction errors.}
\label{S-fig:T2_us}
\end{figure*}

We then questioned whether the performance benefits of PSFNet are also apparent during unsupervised training of deep network models. For this purpose, unsupervised variants PSFNet\SB{US} and MoDL\SB{US} were trained via self-supervision \cite{yaman2020}. PSFNet\SB{US} was compared against MoDL\SB{US}, SPIRiT, sRAKI-RNN, and SPARK while the number of training samples ($N_{samples}$) was systematically varied. Figure \ref{S-fig:plot_subs_us} displays PSNR performance for cT\textsubscript{1}-weighted and T\textsubscript{2}-weighted image reconstruction as a function of $N_{samples}$. Similar to the supervised setting, PSFNet\SB{US} outperforms MoDL\SB{US} for all values of $N_{samples}$ ($p<0.05$), and the performance benefits are more noticeable at lower $N_{samples}$. In this case, however, MoDL\SB{US} is unable to reach the performance of the best performing SS method (SPARK) even at $N_{samples}=500$. In contrast, PSFNet\SB{US} starts outperforming SPARK with approximately $N_{samples}=50$ (5 subjects). The enhanced reconstruction quality with PSFNet\SB{US} is corroborated in representative reconstructions for cT\textsubscript{1}- and T\textsubscript{2}-weighted images depicted in Figures \ref{S-fig:T1c_us} and \ref{S-fig:T2_us}, where $N_{samples}=100$ were used for training. Taken together, these results indicate that the data-efficient nature of PSFNet facilitates the training of both supervised and unsupervised MRI reconstruction models.

\subsection{Generalization Performance}
An important advantage of SS priors is that they allow model adaptation to individual test samples, thereby promise enhanced performance in out-of-domain reconstructions \cite{Knoll2019inverseGANs}. Yet, SG priors with fixed parameters might show relatively limited generalizability during inference \cite{Konukoglu2019,dalmaz2022one}. To assess generalization performance, we introduced domain variations by altering three experimental factors: tissue contrast, undersampling pattern, and acceleration rate. For methods comprising SG components, we built both target-domain models that were trained in the target domain, and domain-transferred models that were trained in a non-target domain. We then compared the reconstruction performances of the two models in the target domain. 

First, we examined generalization performance when the tissue contrast varied between training and testing domains (e.g., trained on cT\SB{1}, tested on T\SB{2}). Table \ref{tab:generalization_contrast} lists performance metrics for competing methods with $N_{samples}=500$. While performance losses are incurred for domain-transferred PSFNet-DT and MoDL-DT models that contain SG components, these losses are modest. On average, MoDL-DT shows a loss of 0.3dB PSNR and 0.1\% SSIM ($p<0.05$), and PSFNet-DT shows a loss of 0.2dB PSNR and 0.1\% SSIM ($p<0.05$). Note that PSFNet-DT still outperforms the closest competing SS method by 2.2dB PSNR and 1.8\% SSIM ($p<0.05$).

Second, we examined generalization performance when models were trained with variable-density and tested on uniform-density undersampling patterns. Table \ref{tab:generalization_mask} lists performance metrics for competing methods. On average across tissue contrasts, MoDL-DT suffers a notable performance loss of 3.6dB PSNR and 2.5\% SSIM ($p<0.05$). In contrast, PSFNet-DT shows a relatively limited loss of 0.4dB PSNR and 0.2\% SSIM ($p<0.05$). Note that PSFNet-DT again outperforms the closest competing SS method by 3.4dB PSNR and 3.7\% SSIM ($p<0.05$).

Third, we examined generalization performance when models were trained at R=4x and tested on R=8x. Table \ref{tab:generalization_R} lists performance metrics for competing methods. On average across tissue contrasts, MoDL-DT suffers a notable performance loss of 1.0dB PSNR and performs slightly better in SSIM by 0.2\%SSIM ($p<0.05$), whereas PSFNet-DT shows a lower loss of 0.6dB PSNR ($p<0.05$) and performs similarly in SSIM ($p>0.05$). PSFNet-DT outperforms the closest competing SS method by 1.2dB PSNR and 1.9\% SSIM ($p<0.05$). Taken together, these results clearly suggest that the SS prior in PSFNet contributes to its improved generalization performance over the scan-general MoDL method, while the SG prior in PSFNet enables it to outperform competing SS methods.  

\begin{table}[t]
\centering
\captionsetup{justification=justified ,width=\linewidth}
\caption{Generalization across tissue contrasts. PSNR and SSIM values (mean$\pm$standard error) across test subjects. Results are shown for scan-specific models (SPIRiT, SPARK, sRAKI-RNN), target-domain models (MoDL, PSFNet) and domain-transferred models (MoDL-DT, PSFNet-DT) at R=4x. The tissue contrast in the target domain is listed in the left-most column (cT\SB{1} or T\SB{2}), domain-transferred models were trained for the non-target tissue contrast. 
}
\begin{center}
\resizebox{\columnwidth}{!}
{%
\begin{tabular}{|*{8}{>{\centering\arraybackslash}p{.15\columnwidth}|}}
\hline
&
{\textbf{SPIRiT}}   &
{\textbf{SPARK}}   &
 {\textbf{sRAKI-RNN}}   &
{\textbf{MoDL}} &
  {\textbf{MoDL-DT}} &
{\textbf{PSFNet}} &
   {\textbf{PSFNet-DT}}\\  
    
 \hline
 & 
  \multicolumn{7}{c|}{PSNR} 
\\ \hline
   \multirow{2}{*}{cT\SB{1}} &  37.6  & 37.6 & 36.8 &  38.5 & 38.2  & 39.9 & 39.4  \\ 
&  $\pm$1.5 &   $\pm$1.5 &  $\pm$1.3  &  $\pm$ 1.5 &  $\pm$1.5 &  $\pm$1.7 &  $\pm$1.6 \\ \hline
\multirow{2}{*}{T\SB{2}}  &  35.8  & 36.5 & 35.2 &  37.9 & 37.5  & 39.0 & 39.0  \\ 
&  $\pm$1.0 &   $\pm$1.0 &  $\pm$1.1  &  $\pm$ 1.0 &  $\pm$1.1 &  $\pm$1.0 &  $\pm$0.9 \\\hline
& \multicolumn{7}{c|}{SSIM} 
\\ \hline
\multirow{2}{*}{cT\SB{1}} &  93.1  & 93.3 & 93.8 &  95.1 & 94.8  & 95.8 & 95.6 \\ 
&  $\pm$1.5  &  $\pm$1.4 &  $\pm$1.0   &   $\pm$1.0 &  $\pm$1.1 &  $\pm$1.0  &  $\pm$1.0  \\ \hline
\multirow{2}{*}{T\SB{2}} &  90.8  & 93.1 & 94.9 &  96.2 & 96.2  & 96.7 & 96.8 \\ 
&  $\pm$1.2  &  $\pm$1.0 &  $\pm$0.6   &   $\pm$0.5 &  $\pm$0.5 &  $\pm$0.4  &  $\pm$0.4  \\ \hline
\end{tabular}%
}
\end{center}%
\label{tab:generalization_contrast}
\end{table}

\begin{table}[t]
\centering
\captionsetup{justification=justified,width=\linewidth}
\caption{Generalization across undersampling patterns. PSNR and SSIM values (mean$\pm$standard error) across test subjects. Results are shown for Results are shown for scan-specific models (SPIRiT, SPARK, sRAKI-RNN), target-domain models (MoDL, PSFNet) and domain-transferred models (MoDL-DT, PSFNet-DT) at R=4x. Domain-transferred models were trained with variable-density undersampling, and tested on uniform-density undersampling. Target-domain models were trained and tested with uniform-density undersampling.}
\begin{center}
\resizebox{\columnwidth}{!}
{%
\begin{tabular}{|*{8}{>{\centering\arraybackslash}p{.15\columnwidth}|}}
\hline
&
{\textbf{SPIRiT}}   &
{\textbf{SPARK}}   &
 {\textbf{sRAKI-RNN}}   &
{\textbf{MoDL}} &
  {\textbf{MoDL-DT}} &
{\textbf{PSFNet}} &
   {\textbf{PSFNet-DT}}\\  
    
 \hline
 & 
  \multicolumn{7}{c|}{PSNR} 
\\ \hline
   \multirow{2}{*}{cT\SB{1}} &  37.1  & 37.1 & 33.6 &  37.0 & 33.6  & 40.2 & 39.9  \\ 
&  $\pm$1.8 &   $\pm$1.7 &  $\pm$1.4  &  $\pm$ 1.7 &  $\pm$1.8 &  $\pm$1.6 &  $\pm$1.6 \\ \hline
\multirow{2}{*}{T\SB{2}}  &  35.1  & 35.6 & 31.6 &  37.0 & 33.2  & 40.2 & 39.7  \\ 
&  $\pm$1.3 &   $\pm$1.3 &  $\pm$1.5  &  $\pm$ 1.1 &  $\pm$1.2 &  $\pm$1.1 &  $\pm$1.2 \\\hline
& \multicolumn{7}{c|}{SSIM} 
\\ \hline
\multirow{2}{*}{cT\SB{1}} &  92.9  & 93.0 & 91.2 &  93.4 & 91.2  & 95.9 & 95.6 \\ 
&  $\pm$1.5  &  $\pm$1.5 &  $\pm$1.5   &   $\pm$1.3 &  $\pm$2.0 &  $\pm$1.2  &  $\pm$1.2  \\ \hline
\multirow{2}{*}{T\SB{2}} &  90.6  & 92.1 & 91.5 &  95.6 & 92.7  & 97.1 & 96.9 \\ 
&  $\pm$1.5  &  $\pm$1.5 &  $\pm$1.2   &   $\pm$0.7 &  $\pm$1.1 &  $\pm$0.6  &  $\pm$0.6  \\ \hline
\end{tabular}%
}
\end{center}%

\label{tab:generalization_mask}
\end{table}

\begin{table}[t]
\centering
\captionsetup{justification=justified ,width=\linewidth}
\caption{Generalization across acceleration rates. PSNR and SSIM values (mean$\pm$standard error) across test subjects. Results are shown for scan-specific models (SPIRiT, SPARK, sRAKI-RNN), target-domain models (MoDL, PSFNet) and domain-transferred models (MoDL-DT, PSFNet-DT). Domain-transferred models were trained at R=4x and tested at R=8x. Target-domain models were trained and tested at R=8x.
}
\begin{center}
\resizebox{\columnwidth}{!}
{%
\begin{tabular}{|*{8}{>{\centering\arraybackslash}p{.15\columnwidth}|}}
\hline
&
{\textbf{SPIRiT}}   &
{\textbf{SPARK}}   &
 {\textbf{sRAKI-RNN}}   &
{\textbf{MoDL}} &
  {\textbf{MoDL-DT}} &
{\textbf{PSFNet}} &
   {\textbf{PSFNet-DT}}\\  
    
 \hline
 & 
  \multicolumn{7}{c|}{PSNR} 
\\ \hline
   \multirow{2}{*}{cT\SB{1}} &  34.7  & 34.8 & 34.3 &  35.3 & 34.5  & 36.5 & 36.2  \\ 
&  $\pm$1.5 &   $\pm$1.5 &  $\pm$1.5  &  $\pm$ 1.4 &  $\pm$1.7 &  $\pm$1.5 &  $\pm$1.5 \\ \hline
\multirow{2}{*}{T\SB{2}}  &  33.6  & 33.7 & 32.6 &  34.6 & 33.4  & 35.6 & 34.6  \\ 
&  $\pm$1.0 &   $\pm$1.0 &  $\pm$0.9  &  $\pm$ 1.0 &  $\pm$1.2 &  $\pm$1.1 &  $\pm$1.2 \\\hline
& \multicolumn{7}{c|}{SSIM} 
\\ \hline
\multirow{2}{*}{cT\SB{1}} &  89.8  & 90.8 & 91.4 &  92.1 & 92.2  & 93.3 & 93.3 \\ 
&  $\pm$1.9  &  $\pm$1.6 &  $\pm$1.4   &   $\pm$1.5 &  $\pm$1.4 &  $\pm$1.4  &  $\pm$1.4  \\ \hline
\multirow{2}{*}{T\SB{2}} &  89.0  & 90.1 & 92.7 &  93.5 & 93.7  & 94.6 & 94.5 \\ 
&  $\pm$1.3  &  $\pm$1.1 &  $\pm$0.9   &   $\pm$0.8 &  $\pm$0.8 &  $\pm$0.7  &  $\pm$0.7  \\ \hline
\end{tabular}%
}
\end{center}%

\label{tab:generalization_R}
\end{table}

\subsection{Sensitivity to Hyperparameters}
Parameters of deep networks that implement SS priors are to be learned from a single test sample, so the resultant models can show elevated sensitivity to the selection of hyperparameters compared to SG priors learned from a collection of training samples. Thus, we investigated the sensitivity of PSFNet to key hyperparameters of its SS prior. SPIRiT, SPARK and PSFNet methods all embody a linear k-space reconstruction, so the relevant hyperparameters are the regularization weight and width for the convolution kernel. Performance was evaluated for models were trained in the low-data regime (i.e., $N_{samples}=10$, 1 subject) for varying hyperparameter values.

\begin{figure}[t]
\centering
\includegraphics[width=0.8\linewidth]{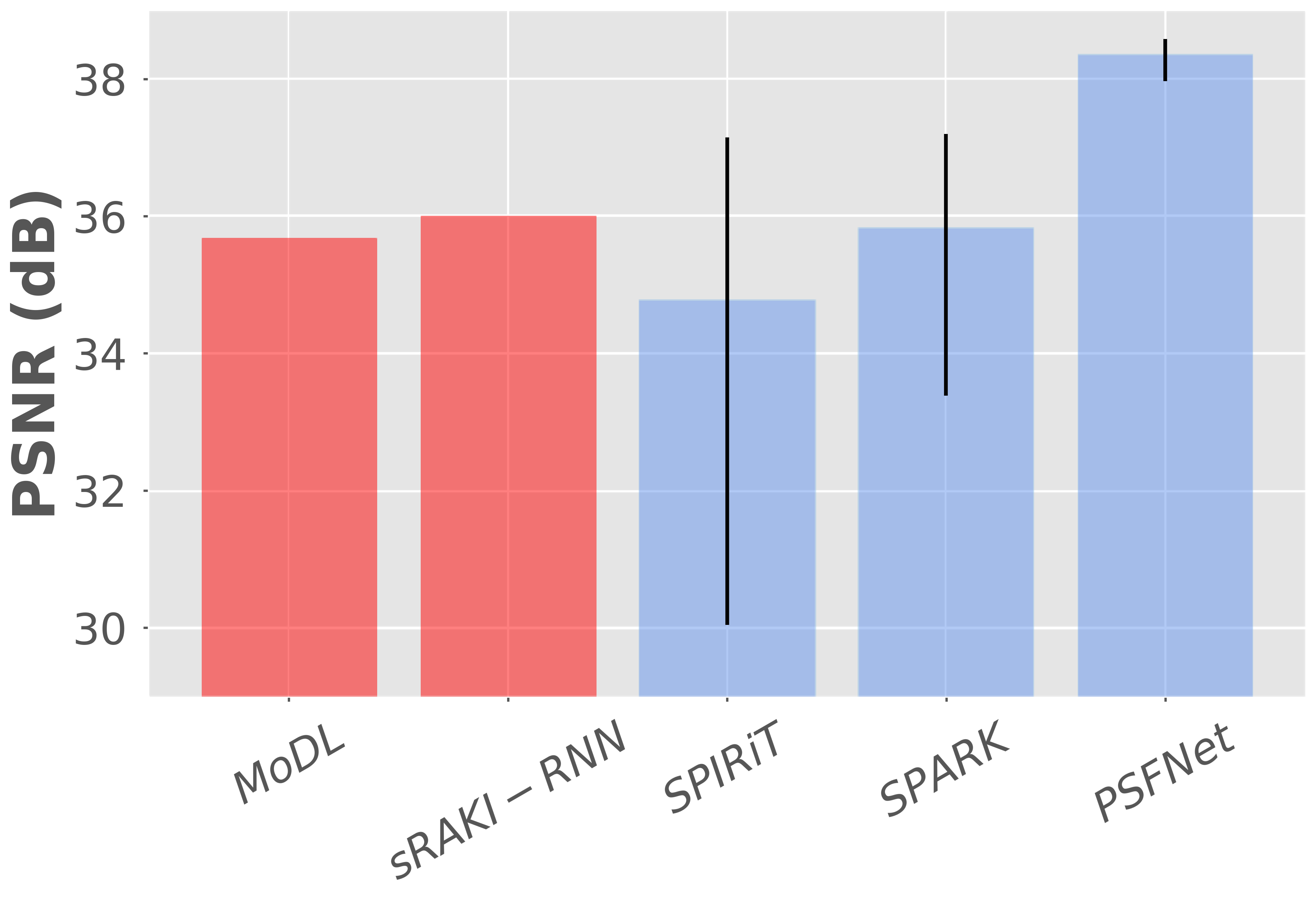}
\caption{PSNR measurements were performed on recovered cT\textsubscript{1}- and T\textsubscript{2}-weighted images at R=4x. Bar plots in blue color show average PSNR across $\kappa \in$ 10\textsuperscript{-3}-10\textsuperscript{1} (i.e., the regularization parameter for kernel estimation). Error bars denote the 90\% interval across $\kappa$. Bar plots in red color show PSNR for methods that do not depend on the value of $\kappa$.}
\label{S-fig:plot_tikhonov}
\end{figure}

\begin{figure}[t] 
\centering
\includegraphics[width=0.8\linewidth]{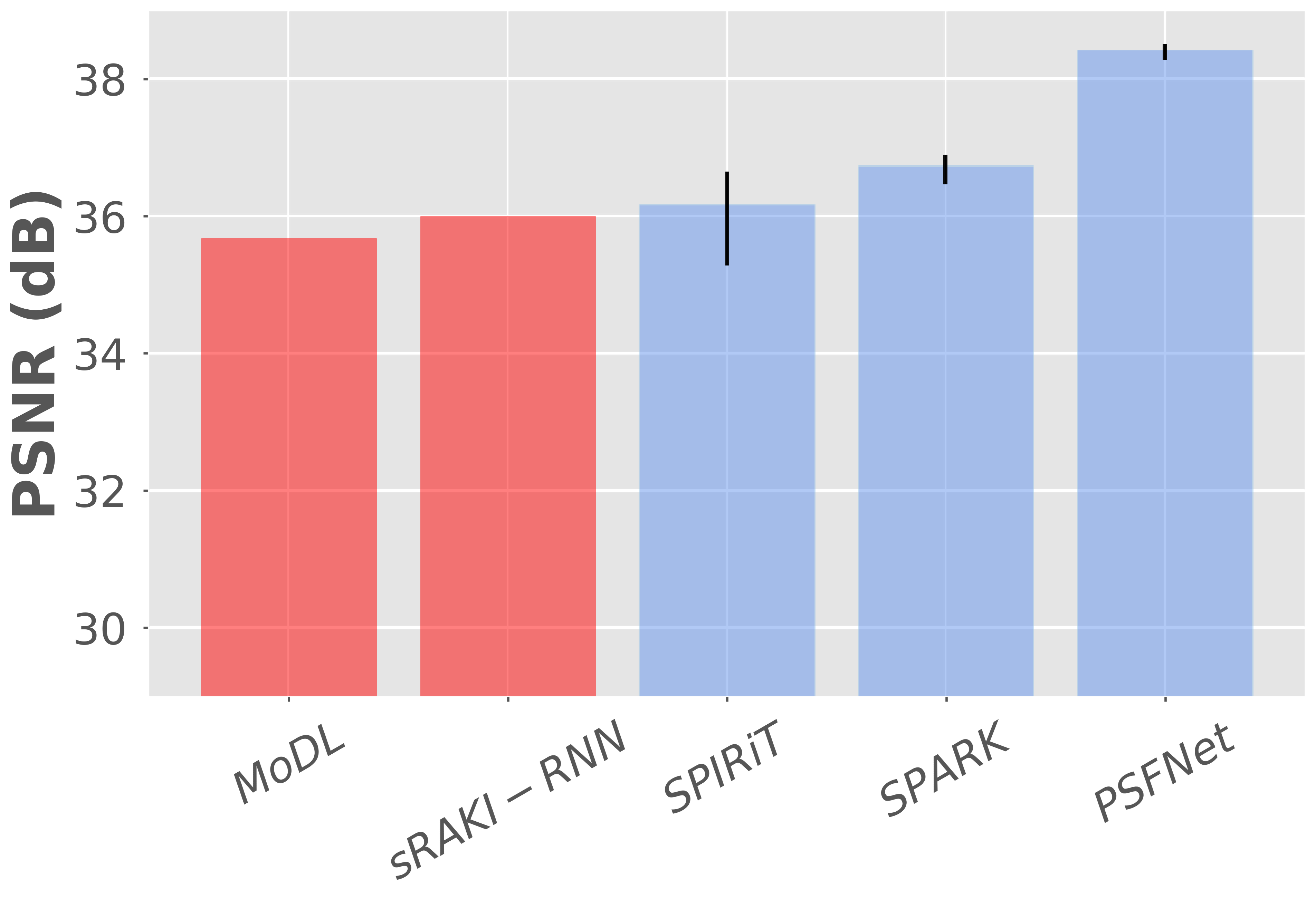}
\caption{PSNR measurements were performed on recovered cT\textsubscript{1}- and T\textsubscript{2}-weighted images at R=4x. Bar plots in blue color show the average PSNR across $w \in$ 5-17 (i.e., the kernel size). Error bars denote the 90\% interval across $w$. Bar plots in red color show PSNR for methods that do not depend on the value of $w$.}
\label{S-fig:plot_kernel}
\end{figure}

Figure \ref{S-fig:plot_tikhonov} displays PSNR measurements for SPIRiT, SPARK and PSFNet across $\kappa$ in range (10\textsuperscript{-3}-10\textsuperscript{0}). While the performance of SPIRiT and SPARK is notably influenced by $\kappa$, PSFNet is minimally affected by sub-optimal selection. On average across contrasts, the difference between the maximum and minimum PSNR values is 8.4dB for SPIRiT, 4.5dB for SPARK, and a lower 0.7dB for PSFNet. Note that PSFNet outperforms competing methods across the entire range of $\kappa$ ($p<0.05$). Figure \ref{S-fig:plot_kernel} shows PSNR measurements for competing methods across $w$ in range (5-17). In this case, all methods show relatively limited sensitivity to the selection of $w$. On average across contrasts, the difference between the maximum and minimum PSNR values is 1.5dB for SPIRiT, 0.5dB for SPARK, and 0.2dB for PSFNet. Again, PSFNet outperforms competing methods across the entire range of $w$ ($p<0.05$). Overall, our results indicate that PSFNet yields improved reliability against sub-optimal hyperparameter selection than competing SS methods.

\subsection{Computational Complexity}
Next, we assessed the computational complexity of competing methods. Table \ref{tab:inference_time} lists the training times of methods with SG priors, MoDL and PSFNet. Note that the remaining SS based methods do not involve a pre-training step. As it involves learning of an SS prior on each training sample, PSFNet yields elevated training time compared to MoDL. In return, however, it offers enhanced generalization performance and data-efficient learning. Table \ref{tab:inference_time} also lists the inference times of SPIRiT, SPARK, sRAKI-RNN, MoDL and PSFNet. MoDL and PSFNet that employ SG priors with fixed weights during inference offer fast run times. In contrast, SPARK and sRAKI-RNN that involve SS priors learned on individual test samples have a high computational burden. Although PSFNet also embodies an SS prior, its uses a relatively lightweight linear prior as opposed to the nonlinear priors in competing SS methods. Therefore, PSFNet benefits from data-efficient learning while maintaining computationally-efficient inference. 

\begin{table}[t]
\centering
\captionsetup{justification=justified ,width=\linewidth}
\caption{Computational complexity of competing methods. Training and inference times for data from a single subject, with 10 cross-sections, imaging matrix size 256x320 and 5 coils. Run times are listed for SPARK, sRAKI-RNN, MoDL, and PSFNet. 
}
\resizebox{\columnwidth}{!}
{
\begin{tabular}{|*{6}{>{\centering\arraybackslash}p{.2\columnwidth}|}}
\hline
&
  %
  \multicolumn{1}{c|}{\textbf{SPIRiT}} &
  \multicolumn{1}{c|}{\textbf{SPARK}} &
  \multicolumn{1}{c|}{\textbf{sRAKI-RNN}} &
  \multicolumn{1}{c|}{\textbf{MoDL}} &
  \multicolumn{1}{c|}{\textbf{PSFNet}}\\ \hline

   \multirow{1}{*}{Training(s)} 
   &   - &   - & - &   132  &337 \\ \hline
   \multirow{1}{*}{Inference(s)} 
   &   0.85 &   23.35  & 285.00 & 0.25 &1.13\\ \hline
\end{tabular}%
}
\label{tab:inference_time}
\end{table}

\subsection{Ablation Analysis}
\begin{figure}[t]
\centering
\includegraphics[width=0.9\linewidth]{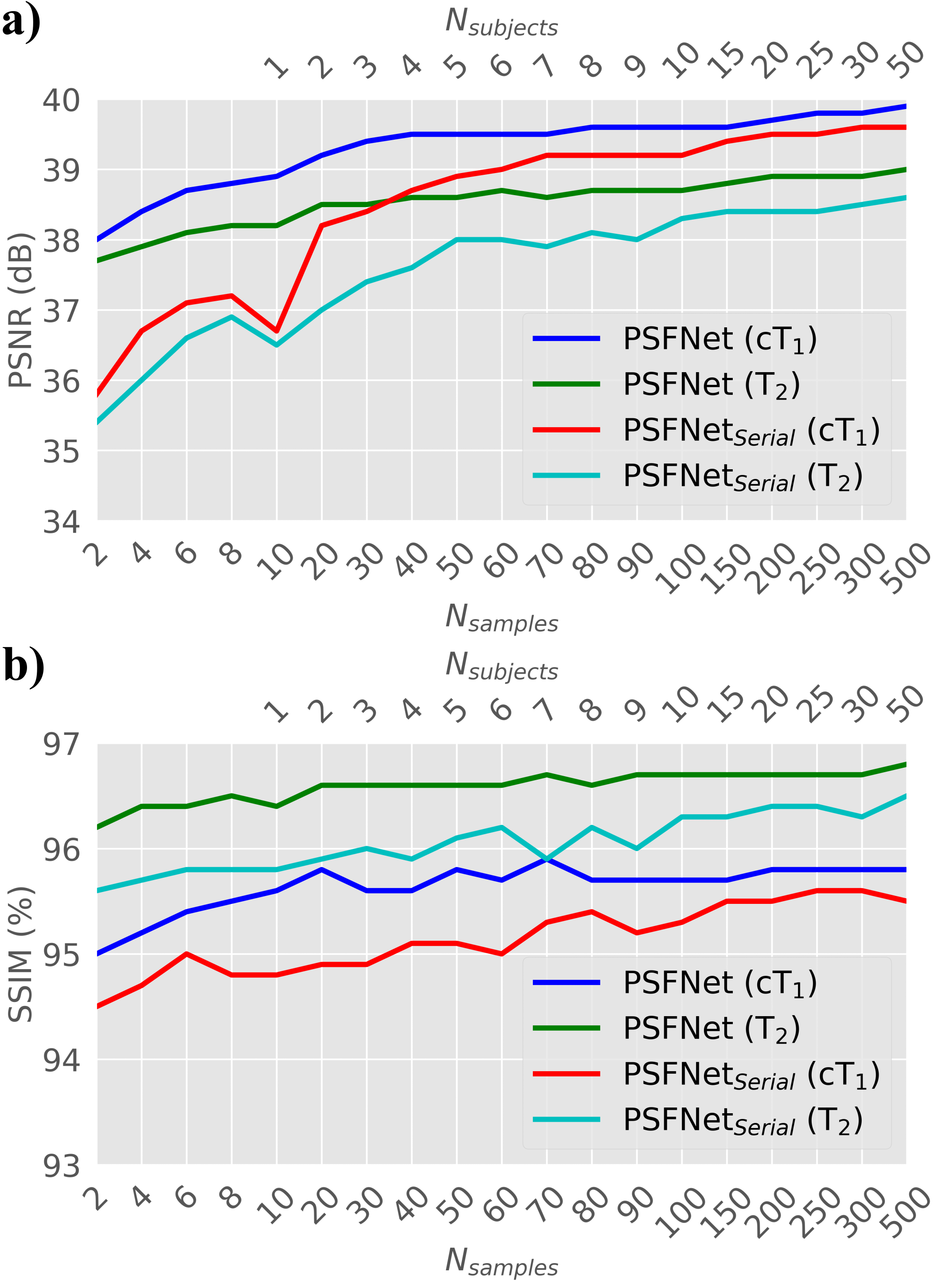}
\caption{Average \textbf{(a)} PSNR and \textbf{(b)} SSIM values for cT\textsubscript{1}- and T\textsubscript{2}-weighted image reconstructions at R=4x. Model training was performed for varying number of training samples ($N_{samples}$, lower x-axis) and thereby training subjects ($N_{subjects}$, upper x-axis). Results are shown for PSFNet and PSFNet\SB{Serial}.}
\label{fig:ablationfigure}
\end{figure}

To demonstrate the value of the parallel-stream fusion strategy in PSFNet over conventional unrolling, PSFNet was compared against a variant model PSFNet\SB{Serial} that combined SS and SG priors through serially alternated projections. Separate models were trained with number of training samples in the range $N_{samples}$=[2-500]. Performance in cT\SB{1}- and T\SB{2} -weighted image reconstruction is displayed in Figure \ref{fig:ablationfigure}. PSFNet significantly improves reconstruction performance over PSFNet\SB{Serial} across the entire range of $N_{samples}$ considered ($p<0.05$), and the benefits grow stronger for smaller training sets. On average across contrasts for $N_{samples} < 10$, PSFNet outperforms PSFNet\SB{Serial} by 1.8dB PSNR and 0.6\% SSIM ($p<0.05$). These results indicate that the parallel-stream fusion of SG and SS priors in PSFNet is superior to the serial projections in conventional unrolling. 

\section{Discussion and Conclusion}
In this study, we introduced PSFNet for data-efficient training of deep reconstruction models in accelerated MRI. PSFNet synergistically fuses SS and SG priors in a parallel-stream architecture. The linear SS prior improves learning efficiency while mataining relatively low computational footprint, whereas the nonlinear SG prior enables improved reconstruction performance. For both supervised and unsupervised training setups, the resulting model substantially reduces dependence on the availability of large MRI datasets. Furthermore, it achieves competitive inference times to SG methods, and reliably generalizes across tissue contrasts, sampling patterns and acceleration rates.

Several prominent approaches have been introduced in the literature to address the training requirements of deep models based on SG priors. One approach is to pre-train models on readily available datasets from a separate source domain and then to fine-tune on several tens of samples from the target domain \cite{Dar2017,KnollGeneralization} or else perform SS fine-tuning \cite{fine_tuning}. This transfer learning approach relaxes the domain requirements for training datasets. However, the domain-transferred models might be suboptimal when training and testing data distributions are divergent. In such cases, additional training for domain-alignment might be necessary to mitigate performance losses. In contrast, PSFNet contains a SS prior that allows it to better generalize to out-of-domain data without further training. Another approach is to build unsupervised models to alleviate dependency on training datasets with paired undersampled, fully-sampled acquisitions. Model training can be performed either directly on undersampled acquisitions via self-supervision \cite{yaman2020} or on unpaired sets of undersampled and fully-sampled acquisitions via cycle-consistent learning \cite{wgan}. This approach can prove beneficial when fully-sampled acquisitions are costly to collect. Nonetheless, the resulting models still require relatively large datasets form tens of subjects during training \cite{yaman2020}. Note that our experiments on self-supervised variants of PSFNet and MoDL suggest that unsupervised models can be more demanding for data than their supervised counterparts. Therefore, the data-efficiency benefits of PSFNet might be particularly useful for unsupervised deep MRI reconstruction.

A fundamentally different framework to lower requirements on training datasets while offering improved generalizability is based on SS priors. In this case, learning can be performed directly on test data and models can be adapted to each scan \cite{raki,spark}. A group of studies have proposed SS methods based on relatively compact nonlinear models to facilitate learning during inference \cite{raki,sraki,rraki,spark}. However, because learning is performed in central k-space, these methods implicitly assume that local relationships among spatial frequency samples are largely invariant across k-space. While the SS prior in PSFNet also rests on a similar assumption, the SG components helps correct residual errors that can be introduced due to this assumption. Another group of studies have alternatively adopted the deep image prior (DIP) approach to build SS methods \cite{Knoll2019inverseGANs,Konukoglu2019,Yilmaz2021,Arora2020ismrm}. In DIP, unconditional deep network models that map latent variables onto images are used as native priors for MR images. The priors are learned by ensuring the consistency of reconstructed and acquired data across the entire k-space. Despite improved generalization, these relatively more complex models require increased inference times. In comparison, PSFNet provides faster inference since the weights for its SG prior are fixed, and its SS prior involves a compact linear operator that is easier to learn.


Few independent studies on MRI have proposed approaches related to PSFNet by combining nonlinear and linear reconstructions \cite{rraki,spark,Grappa_net}. Residual RAKI and SPARK methods initially perform a linear reconstruction, and then use an SS method to correct residual errors via minimizing a DC loss in the calibration region \cite{rraki,spark}. As local relationships among data samples might vary across k-space, the learned SS priors might be suboptimal. Moreover, these methods perform online learning of nonlinear SS priors that introduces relatively high computational burden. In contrast, PSFNet incorporates an SG prior to help improve reliability against sub-optimalities in the SS prior, and uses a linear SS prior for efficiency. Another related method is GrappaNet that improves reconstruction performance by cascading GRAPPA and network-based nonlinear reconstruction steps \cite{Grappa_net}. While \cite{Grappa_net} intends to improve image quality, the main aim of our study is to improve practicality by lowering training data requirements of deep models, and improving domain generalizability without elevating inference times. Note that GrappaNet follows the conventional unrolling approach by performing serially alternated projections through linear and nonlinear reconstructions, which can lead to error propagation under low-data regimes \cite{Murphy2012}. In contrast, PSFNet maintains linear and nonlinear reconstructions as two parallel streams in its architecture, and learns to optimally fuse the information from the two streams. 

The proposed method can be improved along several lines of technical development. First, to improve the capture of high-frequency information by the SG prior, an adversarial loss term along with a discriminator subnetwork can be included in PSFNet \cite{ozbey2022unsupervised}. It remains to be demonstrated whether the data-efficiency benefits of PSFNet are apparent for adversarial training setups. Second, nonlinear activation functions can be included in the SS stream to improve the expressiveness of the SS prior \cite{rraki}. While learning of nonlinear priors can elevate inference complexity, generalization performance might be further improved. Third, the expressiveness of both SS and SG priors might be enhanced by incorporating attention mechanisms as proposed in recent transformer models \cite{dalmaz2021resvit}. Fourth, using multimodal image fusion approaches can improve performance in case of having a repository with multimodal data \cite{fuseZHOU2022184,fuseLIU2023205}. Lastly, the benefits of transfer learning and PSFNet can be aggregated by pre-training the SG prior on natural images to further lower requirements on training data.

\section{Acknowledgments}

This work was supported in part by a TUBA GEBIP 2015 fellowship, by a BAGEP 2017 fellowship, and by a TUBITAK 121E488 grant awarded to T. \c{C}ukur.

\bibliographystyle{IEEETran} 
\bibliography{IEEEabrv,PSFNet}

\begin{thebibliography}{10}
\providecommand{\url}[1]{#1}
\csname url@samestyle\endcsname
\providecommand{\newblock}{\relax}
\providecommand{\bibinfo}[2]{#2}
\providecommand{\BIBentrySTDinterwordspacing}{\spaceskip=0pt\relax}
\providecommand{\BIBentryALTinterwordstretchfactor}{4}
\providecommand{\BIBentryALTinterwordspacing}{\spaceskip=\fontdimen2\font plus
\BIBentryALTinterwordstretchfactor\fontdimen3\font minus
  \fontdimen4\font\relax}
\providecommand{\BIBforeignlanguage}[2]{{%
\expandafter\ifx\csname l@#1\endcsname\relax
\typeout{** WARNING: IEEEtran.bst: No hyphenation pattern has been}%
\typeout{** loaded for the language `#1'. Using the pattern for}%
\typeout{** the default language instead.}%
\else
\language=\csname l@#1\endcsname
\fi
#2}}
\providecommand{\BIBdecl}{\relax}
\BIBdecl

\bibitem{MRI_line}
S.~Bauer, R.~Wiest, L.-P. Nolte, and M.~Reyes, ``A survey of mri-based medical
  image analysis for brain tumor studies,'' \emph{Physics in Medicine \&
  Biology}, vol.~58, no.~13, p. R97, 2013.

\bibitem{SHOEIBI202385}
A.~Shoeibi, M.~Khodatars, M.~Jafari, N.~Ghassemi, P.~Moridian, R.~Alizadehsani,
  S.~H. Ling, A.~Khosravi, H.~Alinejad-Rokny, H.~Lam, M.~Fuller-Tyszkiewicz,
  U.~R. Acharya, D.~Anderson, Y.~Zhang, and J.~M. Gorriz, ``Diagnosis of brain
  diseases in fusion of neuroimaging modalities using deep learning: A
  review,'' \emph{Information Fusion}, vol.~93, pp. 85--117, 2023.

\bibitem{classHU2022330}
M.~Hu, X.~Qian, S.~Liu, A.~J. Koh, K.~Sim, X.~Jiang, C.~Guan, and J.~H. Zhou,
  ``Structural and diffusion mri based schizophrenia classification using 2d
  pretrained and 3d naive convolutional neural networks,'' \emph{Schizophrenia
  Research}, vol. 243, pp. 330--341, 2022.

\bibitem{segmenFERNANDO2023450}
K.~R.~M. Fernando and C.~P. Tsokos, ``Deep and statistical learning in
  biomedical imaging: State of the art in 3d mri brain tumor segmentation,''
  \emph{Information Fusion}, vol.~92, pp. 450--465, 2023.

\bibitem{segmZHU2023376}
Z.~Zhu, X.~He, G.~Qi, Y.~Li, B.~Cong, and Y.~Liu, ``Brain tumor segmentation
  based on the fusion of deep semantics and edge information in multimodal
  mri,'' \emph{Information Fusion}, vol.~91, pp. 376--387, 2023.

\bibitem{Grappa_net}
A.~Sriram, J.~Zbontar, T.~Murrell, C.~L. Zitnick, A.~Defazio, and D.~K.
  Sodickson, ``{GrappaNet}: {Combining} parallel imaging with deep learning for
  multi-coil {MRI} reconstruction,'' in \emph{Proceedings of the IEEE/CVF
  Conference on Computer Vision and Pattern Recognition (CVPR)}, June 2020, pp.
  14\,303--14\,310.

\bibitem{Pruessmann1999}
K.~P. Pruessmann, M.~Weiger, M.~B. Scheidegger, and P.~Boesiger, ``{SENSE:
  sensitivity encoding for fast MRI.}'' \emph{Magnetic Resonance in Medicine},
  vol.~42, no.~5, pp. 952--62, 1999.

\bibitem{Griswold2002}
M.~A. Griswold, P.~M. Jakob, R.~M. Heidemann, M.~Nittka, V.~Jellus, J.~Wang,
  B.~Kiefer, and A.~Haase, ``{Generalized autocalibrating partially parallel
  acquisitions (GRAPPA)},'' \emph{Magnetic Resonance in Medicine}, vol.~47,
  no.~6, pp. 1202--1210, 2002.

\bibitem{Lustig2007}
M.~Lustig, D.~Donoho, and J.~M. Pauly, ``{Sparse MRI: The application of
  compressed sensing for rapid MR imaging},'' \emph{Magnetic Resonance in
  Medicine}, vol.~58, no.~6, pp. 1182--1195, 2007.

\bibitem{majumdar2015improving}
A.~Majumdar, ``Improving synthesis and analysis prior blind compressed sensing
  with low-rank constraints for dynamic mri reconstruction,'' \emph{Magnetic
  resonance imaging}, vol.~33, no.~1, pp. 174--179, 2015.

\bibitem{Lustig2010}
M.~Lustig and J.~M. Pauly, ``{SPIRiT: Iterative self-consistent parallel
  imaging reconstruction from arbitrary k-space.}'' \emph{Magnetic Resonance in
  Medicine}, vol.~64, no.~2, pp. 457--71, 2010.

\bibitem{ADMM-CSNET}
Y.~{Yang}, J.~{Sun}, H.~{Li}, and Z.~{Xu}, ``{ADMM-CSNet: A} deep learning
  approach for image compressive sensing,'' \emph{IEEE Transactions on Pattern
  Analysis and Machine Intelligence}, vol.~42, no.~3, pp. 521--538, 2020.

\bibitem{Schlemper2017}
J.~Schlemper, J.~Caballero, J.~V. Hajnal, A.~Price, and D.~Rueckert, ``{A Deep
  Cascade of Convolutional Neural Networks for {MR} Image Reconstruction},'' in
  \emph{International Conference on Information Processing in Medical Imaging},
  2017, pp. 647--658.

\bibitem{Hammernik2017}
K.~Hammernik, T.~Klatzer, E.~Kobler, M.~P. Recht, D.~K. Sodickson, T.~Pock, and
  F.~Knoll, ``Learning a variational network for reconstruction of accelerated
  {MRI} data,'' \emph{Magnetic Resonance in Medicine}, vol.~79, no.~6, pp.
  3055--3071, 2017.

\bibitem{raki}
M.~Akçakaya, S.~Moeller, S.~Weingärtner, and K.~Uğurbil, ``Scan-specific
  robust artificial-neural-networks for k-space interpolation {(RAKI)}
  reconstruction: {Database}-free deep learning for fast imaging,''
  \emph{Magnetic Resonance in Medicine}, vol.~81, no.~1, pp. 439--453, 2019.

\bibitem{loraki}
T.~H. Kim, P.~Garg, and J.~P. Haldar, ``{LORAKI}: Autocalibrated recurrent
  neural networks for autoregressive {MRI} reconstruction in k-space,''
  \emph{arXiv preprint arXiv:1904.09390}, 2019.

\bibitem{spark}
Y.~Arefeen, O.~Beker, J.~Cho, H.~Yu, E.~Adalsteinsson, and B.~Bilgic,
  ``Scan-specific artifact reduction in k-space (spark) neural networks
  synergize with physics-based reconstruction to accelerate mri,''
  \emph{Magnetic Resonance in Medicine}, vol.~87, no.~2, pp. 764--780, 2022.

\bibitem{sraki}
S.~A.~H. Hosseini, C.~Zhang, S.~Weingärtner, S.~Moeller, M.~Stuber,
  K.~Ugurbil, and M.~Akçakaya, ``Accelerated coronary {MRI} with {sRAKI: A}
  database-free self-consistent neural network k-space reconstruction for
  arbitrary undersampling,'' \emph{PLOS ONE}, vol.~15, no.~2, pp. 1--13, 2020.

\bibitem{Yilmaz2021}
Y.~Korkmaz, S.~U. Dar, M.~Yurt, M.~{\"O}zbey, and T.~Cukur, ``Unsupervised mri
  reconstruction via zero-shot learned adversarial transformers,'' \emph{IEEE
  Transactions on Medical Imaging}, vol.~41, no.~7, pp. 1747--1763, 2022.

\bibitem{Arora2020ismrm}
S.~Arora, V.~Roeloffs, and M.~Lustig, ``Untrained modified deep decoder for
  joint denoising and parallel imaging reconstruction,'' in \emph{Proceedings
  of the 28th Annual Meeting of the ISMRM}, 2020, p. 3585.

\bibitem{Darestani2021}
M.~Z. Darestani and R.~Heckel, ``Accelerated mri with un-trained neural
  networks,'' \emph{IEEE Transactions on Computational Imaging}, vol.~7, pp.
  724--733, 2021.

\bibitem{Knoll2019inverseGANs}
D.~Narnhofer, K.~Hammernik, F.~Knoll, and T.~Pock, ``{Inverse GANs for
  accelerated MRI reconstruction},'' in \emph{Proceedings of the SPIE}, vol.
  11138, 2019, pp. 381 -- 392.

\bibitem{Konukoglu2019}
K.~C. {Tezcan}, C.~F. {Baumgartner}, R.~{Luechinger}, K.~P. {Pruessmann}, and
  E.~{Konukoglu}, ``{MR} image reconstruction using deep density priors,''
  \emph{IEEE Transactions on Medical Imaging}, vol.~38, no.~7, pp. 1633--1642,
  2019.

\bibitem{Liu2020mrm}
Q.~Liu, Q.~Yang, H.~Cheng, S.~Wang, M.~Zhang, and D.~Liang, ``Highly
  undersampled magnetic resonance imaging reconstruction using autoencoding
  priors,'' \emph{Magnetic Resonance in Medicine}, vol.~83, no.~1, pp.
  322--336, 2020.

\bibitem{KikiNet}
T.~Eo, Y.~Jun, T.~Kim, J.~Jang, H.-J. Lee, and D.~Hwang, ``{KIKI-net:}
  cross-domain convolutional neural networks for reconstructing undersampled
  magnetic resonance images,'' \emph{Magnetic Resonance in Medicine}, vol.~80,
  no.~5, pp. 2188--2201, 2018.

\bibitem{Mardani2019b}
M.~Mardani, E.~Gong, J.~Y. Cheng, S.~Vasanawala, G.~Zaharchuk, L.~Xing, and
  J.~M. Pauly, ``{Deep generative adversarial neural networks for compressive
  sensing MRI},'' \emph{IEEE Transactions on Medical Imaging}, vol.~38, no.~1,
  pp. 167--179, 2019.

\bibitem{MoDl}
H.~K. {Aggarwal}, M.~P. {Mani}, and M.~{Jacob}, ``{MoDL: Model-Based} deep
  learning architecture for inverse problems,'' \emph{IEEE Transactions on
  Medical Imaging}, vol.~38, no.~2, pp. 394--405, 2019.

\bibitem{Dar2017}
S.~U.~H. Dar, M.~{\"{O}}zbey, A.~B. {\c{C}}atlı, and T.~{\c{C}}ukur, ``A
  transfer-learning approach for accelerated {MRI} using deep neural
  networks,'' \emph{Magnetic Resonance in Medicine}, vol.~84, no.~2, pp.
  663--685, 2020.

\bibitem{lee2018deep}
D.~Lee, J.~Yoo, S.~Tak, and J.~C. Ye, ``Deep residual learning for accelerated
  {MRI} using magnitude and phase networks,'' \emph{IEEE Transactions on
  Biomedical Engineering}, vol.~65, no.~9, pp. 1985--1995, 2018.

\bibitem{guo2021over}
P.~Guo, J.~M.~J. Valanarasu, P.~Wang, J.~Zhou, S.~Jiang, and V.~M. Patel,
  ``Over-and-under complete convolutional rnn for mri reconstruction,'' in
  \emph{International Conference on Medical Image Computing and
  Computer-Assisted Intervention}.\hskip 1em plus 0.5em minus 0.4em\relax
  Springer, 2021, pp. 13--23.

\bibitem{yiasemis2022recurrent}
G.~Yiasemis, J.-J. Sonke, C.~S{\'a}nchez, and J.~Teuwen, ``Recurrent
  variational network: A deep learning inverse problem solver applied to the
  task of accelerated mri reconstruction,'' in \emph{Proceedings of the
  IEEE/CVF Conference on Computer Vision and Pattern Recognition}, 2022, pp.
  732--741.

\bibitem{hou2022idpcnn}
R.~Hou and F.~Li, ``Idpcnn: Iterative denoising and projecting cnn for mri
  reconstruction,'' \emph{Journal of Computational and Applied Mathematics},
  vol. 406, p. 113973, 2022.

\bibitem{ramzi2022nc}
Z.~Ramzi, G.~Chaithya, J.-L. Starck, and P.~Ciuciu, ``Nc-pdnet: a
  density-compensated unrolled network for 2d and 3d non-cartesian mri
  reconstruction,'' \emph{IEEE Transactions on Medical Imaging}, 2022.

\bibitem{Kwon2017}
K.~Kwon, D.~Kim, and H.~Park, ``{A parallel MR imaging method using multilayer
  perceptron},'' \emph{Medical Physics}, vol.~44, no.~12, pp. 6209--6224, 2017.

\bibitem{Wang2016}
S.~Wang, Z.~Su, L.~Ying, X.~Peng, S.~Zhu, F.~Liang, D.~Feng, and D.~Liang,
  ``{Accelerating magnetic resonance imaging via deep learning},'' in
  \emph{IEEE 13th International Symposium on Biomedical Imaging (ISBI)}, 2016,
  pp. 514--517.

\bibitem{ChulYe2018}
J.~C. Ye, Y.~Han, and E.~Cha, ``Deep convolutional framelets: {A} general deep
  learning framework for inverse problems,'' \emph{SIAM Journal on Imaging
  Sciences}, vol.~11, no.~2, pp. 991--1048, 2018.

\bibitem{Yoon2018}
J.~Yoon, E.~Gong, I.~Chatnuntawech, B.~Bilgic, J.~Lee, W.~Jung, J.~Ko, H.~Jung,
  K.~Setsompop, G.~Zaharchuk, E.~Y. Kim, J.~Pauly, and J.~Lee, ``Quantitative
  susceptibility mapping using deep neural network: {QSMnet},''
  \emph{NeuroImage}, vol. 179, pp. 199--206, 2018.

\bibitem{Hyun2018}
C.~M. Hyun, H.~P. Kim, S.~M. Lee, S.~Lee, and J.~K. Seo, ``{Deep learning for
  undersampled MRI reconstruction},'' \emph{Physics in Medicine and Biology},
  vol.~63, no.~13, p. 135007, 2018.

\bibitem{Hauptmann2018}
A.~Hauptmann, S.~Arridge, F.~Lucka, V.~Muthurangu, and J.~A. Steeden,
  ``Real-time cardiovascular {MR} with spatio-temporal artifact suppression
  using deep learning–proof of concept in congenital heart disease,''
  \emph{Magnetic Resonance in Medicine}, vol.~81, no.~2, pp. 1143--1156, 2019.

\bibitem{Hosseini2020b}
S.~A.~H. {Hosseini}, B.~{Yaman}, S.~{Moeller}, M.~{Hong}, and M.~{Akçakaya},
  ``Dense recurrent neural networks for accelerated {MRI: History}-cognizant
  unrolling of optimization algorithms,'' \emph{IEEE Journal of Selected Topics
  in Signal Processing}, vol.~14, no.~6, pp. 1280--1291, 2020.

\bibitem{Conv_recur}
C.~{Qin}, J.~{Schlemper}, J.~{Caballero}, A.~N. {Price}, J.~V. {Hajnal}, and
  D.~{Rueckert}, ``Convolutional recurrent neural networks for dynamic {MR}
  image reconstruction,'' \emph{IEEE Transactions on Medical Imaging}, vol.~38,
  no.~1, pp. 280--290, 2019.

\bibitem{Quan2018c}
T.~M. Quan, T.~Nguyen-Duc, and W.-K. Jeong, ``{Compressed sensing MRI
  reconstruction with cyclic loss in generative adversarial networks},''
  \emph{IEEE Transactions on Medical Imaging}, vol.~37, no.~6, pp. 1488--1497,
  2018.

\bibitem{rgan}
S.~U. Dar, M.~Yurt, M.~Shahdloo, M.~E. Ild{\i}z, B.~T{\i}naz, and
  T.~{\c{C}}ukur, ``Prior-guided image reconstruction for accelerated
  multi-contrast {MRI} via generative adversarial networks,'' \emph{IEEE
  Journal of Selected Topics in Signal Processing}, vol.~14, no.~6, pp.
  1072--1087, 2020.

\bibitem{Chen2021}
Y.~Chen, D.~Firmin, and G.~Yang, ``Wavelet improved {GAN for MRI}
  reconstruction,'' in \emph{Proceedings of SPIE, Medical Imaging 2021: Physics
  of Medical Imaging}, vol. 11595, 2021, p. 1159513.

\bibitem{elmas2022federated}
G.~Elmas, S.~U. Dar, Y.~Korkmaz, E.~Ceyani, B.~Susam, M.~Ozbey, S.~Avestimehr,
  and T.~{\c{C}}ukur, ``Federated learning of generative image priors for mri
  reconstruction,'' \emph{IEEE Transactions on Medical Imaging}, 2022.

\bibitem{yaqub2022gan}
M.~Yaqub, F.~Jinchao, S.~Ahmed, K.~Arshid, M.~A. Bilal, M.~P. Akhter, and M.~S.
  Zia, ``Gan-tl: Generative adversarial networks with transfer learning for mri
  reconstruction,'' \emph{Applied Sciences}, vol.~12, no.~17, p. 8841, 2022.

\bibitem{korkmaz2022mri}
Y.~Korkmaz, M.~{\"O}zbey, and T.~Cukur, ``Mri reconstruction with conditional
  adversarial transformers,'' in \emph{International Workshop on Machine
  Learning for Medical Image Reconstruction}.\hskip 1em plus 0.5em minus
  0.4em\relax Springer, 2022, pp. 62--71.

\bibitem{guo2022reconformer}
P.~Guo, Y.~Mei, J.~Zhou, S.~Jiang, and V.~M. Patel, ``Reconformer: Accelerated
  mri reconstruction using recurrent transformer,'' \emph{arXiv preprint
  arXiv:2201.09376}, 2022.

\bibitem{dar2022adaptive}
A.~G{\"u}ng{\"o}r, S.~U. Dar, {\c{S}}.~{\"O}zt{\"u}rk, Y.~Korkmaz, G.~Elmas,
  M.~{\"O}zbey, and T.~{\c{C}}ukur, ``Adaptive diffusion priors for accelerated
  mri reconstruction,'' \emph{arXiv:2207.05876}, 2022.

\bibitem{peng2022towards}
C.~Peng, P.~Guo, S.~K. Zhou, V.~M. Patel, and R.~Chellappa, ``Towards
  performant and reliable undersampled mr reconstruction via diffusion model
  sampling,'' in \emph{International Conference on Medical Image Computing and
  Computer-Assisted Intervention}.\hskip 1em plus 0.5em minus 0.4em\relax
  Springer, 2022, pp. 623--633.

\bibitem{wang2022high}
K.~Wang, J.~I. Tamir, A.~De~Goyeneche, U.~Wollner, R.~Brada, S.~X. Yu, and
  M.~Lustig, ``High fidelity deep learning-based mri reconstruction with
  instance-wise discriminative feature matching loss,'' \emph{Magnetic
  Resonance in Medicine}, vol.~88, no.~1, pp. 476--491, 2022.

\bibitem{supervisedunrolledGadjimuradov}
F.~Gadjimuradov, T.~Benkert, M.~D. Nickel, and A.~Maier, ``Robust partial
  fourier reconstruction for diffusion-weighted imaging using a recurrent
  convolutional neural network,'' \emph{Magnetic Resonance in Medicine},
  vol.~87, no.~4, pp. 2018--2033, 2022.

\bibitem{supervisedunrolled9684848}
Z.~Ramzi, C.~G~R, J.-L. Starck, and P.~Ciuciu, ``Nc-pdnet: A
  density-compensated unrolled network for 2d and 3d non-cartesian mri
  reconstruction,'' \emph{IEEE Transactions on Medical Imaging}, vol.~41,
  no.~7, pp. 1625--1638, 2022.

\bibitem{Polakjointvvn2020}
D.~Polak, S.~Cauley, B.~Bilgic, E.~Gong, P.~Bachert, E.~Adalsteinsson, and
  K.~Setsompop, ``Joint multi-contrast variational network reconstruction
  {(jVN)} with application to rapid {2D} and {3D} imaging,'' \emph{Magnetic
  Resonance in Medicine}, vol.~84, no.~3, pp. 1456--1469, 2020.

\bibitem{deepspirit}
J.~Y. Cheng, M.~Mardani, M.~T. Alley, J.~M. Pauly, and S.~S. Vasanawala,
  ``Deepspirit: Generalized parallel imaging using deep convolutional neural
  networks,'' in \emph{Proceedings of the 26th Annual Meeting of the ISMRM},
  2018, p. 0570.

\bibitem{supervisedunrolled9761583}
K.~Pooja, Z.~Ramzi, G.~Chaithya, and P.~Ciuciu, ``Mc-pdnet: Deep unrolled
  neural network for multi-contrast mr image reconstruction from undersampled
  k-space data,'' in \emph{2022 IEEE 19th International Symposium on Biomedical
  Imaging (ISBI)}, 2022, pp. 1--5.

\bibitem{tavaf2021grappa}
N.~Tavaf, A.~Torfi, K.~Ugurbil, and P.-F. Van~de Moortele, ``Grappa-gans for
  parallel mri reconstruction,'' \emph{arXiv preprint arXiv:2101.03135}, 2021.

\bibitem{sandino2021accelerating}
C.~M. Sandino, P.~Lai, S.~S. Vasanawala, and J.~Y. Cheng, ``Accelerating
  cardiac cine mri using a deep learning-based espirit reconstruction,''
  \emph{Magnetic Resonance in Medicine}, vol.~85, no.~1, pp. 152--167, 2021.

\bibitem{KnollGeneralization}
F.~Knoll, K.~Hammernik, E.~Kobler, T.~Pock, M.~P. Recht, and D.~K. Sodickson,
  ``Assessment of the generalization of learned image reconstruction and the
  potential for transfer learning,'' \emph{Magnetic Resonance in Medicine},
  vol.~81, no.~1, pp. 116--128, 2019.

\bibitem{chaudhari2021}
A.~S. Chaudhari, C.~M. Sandino, E.~K. Cole, D.~B. Larson, G.~E. Gold, S.~S.
  Vasanawala, M.~P. Lungren, B.~A. Hargreaves, and C.~P. Langlotz,
  ``Prospective deployment of deep learning in mri: A framework for important
  considerations, challenges, and recommendations for best practices,''
  \emph{Journal of Magnetic Resonance Imaging}, vol.~54, no.~2, pp. 357--371,
  2021.

\bibitem{comnet2021}
S.~U.~H. Dar, M.~Yurt, and T.~\c{C}ukur, ``A few-shot learning approach for
  accelerated mri via fusion of data-driven and subject-driven priors,'' in
  \emph{Proceedings of the 29th Annual Meeting of the ISMRM}, 2021, p. 1949.

\bibitem{Tamir2019}
J.~I. Tamir, S.~X. Yu, and M.~Lustig, ``Unsupervised deep basis pursuit:
  {Learning} reconstruction without ground-truth data,'' in \emph{Proceedings
  of the 27th Annual Meeting of the ISMRM}, 2019, p. 0660.

\bibitem{Cole2021}
E.~K. Cole, F.~Ong, S.~S. Vasanawala, and J.~M. Pauly, ``Fast unsupervised mri
  reconstruction without fully-sampled ground truth data using generative
  adversarial networks,'' in \emph{Proceedings of the IEEE/CVF International
  Conference on Computer Vision (ICCV) Workshops}, October 2021, pp.
  3988--3997.

\bibitem{yaman2020}
B.~Yaman, S.~A.~H. Hosseini, S.~Moeller, J.~Ellermann, K.~U{\u{g}}urbil, and
  M.~Ak{\c{c}}akaya, ``Self-supervised learning of physics-guided
  reconstruction neural networks without fully sampled reference data,''
  \emph{Magnetic resonance in medicine}, vol.~84, no.~6, pp. 3172--3191, 2020.

\bibitem{Liu2020}
J.~{Liu}, Y.~{Sun}, C.~{Eldeniz}, W.~{Gan}, H.~{An}, and U.~S. {Kamilov},
  ``{RARE}: Image reconstruction using deep priors learned without
  groundtruth,'' \emph{IEEE Journal of Selected Topics in Signal Processing},
  vol.~14, no.~6, pp. 1088--1099, 2020.

\bibitem{wang2022}
S.~Wang, R.~Wu, C.~Li, J.~Zou, Z.~Zhang, Q.~Liu, Y.~Xi, and H.~Zheng,
  ``{PARCEL: Physics-based Unsupervised Contrastive Representation Learning for
  Multi-coil MR Imaging},'' \emph{arXiv:2202.01494}, 2022.

\bibitem{sraki_rnn}
S.~A.~H. Hosseini, C.~Zhang, K.~Uǧurbil, S.~Moeller, and M.~Ak{\c{c}}akaya,
  ``sraki-rnn: accelerated mri with scan-specific recurrent neural networks
  using densely connected blocks,'' in \emph{Wavelets and Sparsity XVIII}, vol.
  11138.\hskip 1em plus 0.5em minus 0.4em\relax International Society for
  Optics and Photonics, 2019, p. 111381B.

\bibitem{huang2021dynamic}
Q.~Huang, Y.~Xian, D.~Yang, H.~Qu, J.~Yi, P.~Wu, and D.~N. Metaxas, ``Dynamic
  mri reconstruction with end-to-end motion-guided network,'' \emph{Medical
  Image Analysis}, vol.~68, p. 101901, 2021.

\bibitem{huang2022evaluation}
J.~Huang, S.~Wang, G.~Zhou, W.~Hu, and G.~Yu, ``Evaluation on the
  generalization of a learned convolutional neural network for mri
  reconstruction,'' \emph{Magnetic Resonance Imaging}, vol.~87, pp. 38--46,
  2022.

\bibitem{DIP}
D.~Ulyanov, A.~Vedaldi, and V.~Lempitsky, ``Deep image prior,'' in
  \emph{Proceedings of the IEEE Conference on Computer Vision and Pattern
  Recognition (CVPR)}, 2018, pp. 9446--9454.

\bibitem{Uecker2014}
M.~Uecker, P.~Lai, M.~J. Murphy, P.~Virtue, M.~Elad, J.~M. Pauly, S.~S.
  Vasanawala, and M.~Lustig, ``{ESPIRiT-an eigenvalue approach to
  autocalibrating parallel MRI: Where SENSE meets GRAPPA},'' \emph{Magnetic
  Resonance in Medicine}, vol.~71, no.~3, pp. 990--1001, 2014.

\bibitem{fastmri}
F.~Knoll, J.~Zbontar, A.~Sriram, M.~J. Muckley, M.~Bruno, A.~Defazio,
  M.~Parente, K.~J. Geras, J.~Katsnelson, H.~Chandarana, Z.~Zhang,
  M.~Drozdzalv, A.~Romero, M.~Rabbat, P.~Vincent, J.~Pinkerton, D.~Wang,
  N.~Yakubova, E.~Owens, C.~L. Zitnick, M.~P. Recht, D.~K. Sodickson, and Y.~W.
  Lui, ``{fastMRI: A} publicly available raw k-space and {DICOM} dataset of
  knee images for accelerated {MR} image reconstruction using machine
  learning,'' \emph{Radiology: Artificial Intelligence}, vol.~2, no.~1, p.
  e190007, 2020.

\bibitem{Zhang2013}
T.~Zhang, J.~M. Pauly, S.~S. Vasanawala, and M.~Lustig, ``{Coil compression for
  accelerated imaging with Cartesian sampling.}'' \emph{Magnetic Resonance in
  Medicine}, vol.~69, no.~2, pp. 571--82, 2013.

\bibitem{Kingma2015}
D.~P. Kingma and J.~L. Ba, ``{Adam: a Method for Stochastic Optimization},'' in
  \emph{International Conference on Learning Representations}, 2015.

\bibitem{dalmaz2022one}
O.~Dalmaz, U.~Mirza, G.~Elmas, M.~{\"O}zbey, S.~U. Dar, E.~Ceyani,
  S.~Avestimehr, and T.~{\c{C}}ukur, ``One model to unite them all:
  Personalized federated learning of multi-contrast mri synthesis,''
  \emph{arXiv:2207.06509}, 2022.

\bibitem{fine_tuning}
S.~A.~H. Hosseini, B.~Yaman, S.~Moeller, and M.~Akçakaya, ``High-fidelity
  accelerated mri reconstruction by scan-specific fine-tuning of physics-based
  neural networks,'' in \emph{2020 42nd Annual International Conference of the
  IEEE Engineering in Medicine Biology Society (EMBC)}, 2020, pp. 1481--1484.

\bibitem{wgan}
K.~Lei, M.~Mardani, J.~M. Pauly, and S.~S. Vasanawala, ``Wasserstein {GANs} for
  {MR} imaging: from paired to unpaired training,'' \emph{IEEE transactions on
  medical imaging}, vol.~40, no.~1, pp. 105--115, 2021.

\bibitem{rraki}
C.~Zhang, S.~A.~H. Hosseini, S.~Moeller, S.~Weingärtner, K.~Ugurbil, and
  M.~Akcakaya, ``Scan-specific residual convolutional neural networks for fast
  mri using residual raki,'' in \emph{2019 53rd Asilomar Conference on Signals,
  Systems, and Computers}, 2019, pp. 1476--1480.

\bibitem{Murphy2012}
M.~Murphy, M.~Alley, J.~Demmel, K.~Keutzer, S.~Vasanawala, and M.~Lustig,
  ``{Fast l₁-SPIRiT compressed sensing parallel imaging MRI: scalable
  parallel implementation and clinically feasible runtime.}'' \emph{IEEE
  transactions on medical imaging}, vol.~31, no.~6, pp. 1250--1262, 2012.

\bibitem{ozbey2022unsupervised}
M.~{\"O}zbey, O.~Dalmaz, S.~U. Dar, H.~A. Bedel, {\c{S}}.~{\"O}zturk,
  A.~G{\"u}ng{\"o}r, and T.~{\c{C}}ukur, ``Unsupervised medical image
  translation with adversarial diffusion models,'' \emph{arXiv:2207.08208},
  2022.

\bibitem{dalmaz2021resvit}
O.~Dalmaz, M.~Yurt, and T.~\c{C}ukur, ``{ResViT: Residual} vision transformers
  for multi-modal medical image synthesis,'' \emph{IEEE Transactions on Medical
  Imaging}, vol.~41, no.~7, pp. 2598--2614, 2022.

\bibitem{fuseZHOU2022184}
H.~Zhou, J.~Hou, Y.~Zhang, J.~Ma, and H.~Ling, ``Unified gradient- and
  intensity-discriminator generative adversarial network for image fusion,''
  \emph{Information Fusion}, vol.~88, pp. 184--201, 2022.

\bibitem{fuseLIU2023205}
J.~Liu, R.~Dian, S.~Li, and H.~Liu, ``Sgfusion: A saliency guided deep-learning
  framework for pixel-level image fusion,'' \emph{Information Fusion}, vol.~91,
  pp. 205--214, 2023.

\end{thebibliography}

\end{document}